\documentclass[10pt,pra,aps,amssymb,amsmath,tightenlines,twocolumn]{revtex4-2} 

\usepackage{graphicx,epsfig,xcolor}
\usepackage[T1]{fontenc}
\usepackage[utf8]{inputenc}
\usepackage[english]{babel}
\usepackage{amsmath, bbm, nccmath, MnSymbol}
\usepackage{blkarray}
\usepackage{mathtools}
\usepackage[normalem]{ulem}
\usepackage{amsfonts}
\usepackage{amssymb}
\usepackage{bm}
\usepackage{braket}
\usepackage{dsfont}
\usepackage{float}
\usepackage{color}
\definecolor{ikb}{rgb}{0, 0.184,0.655}

\usepackage[labelformat=parens]{subfig}
\usepackage[justification=raggedright, font=footnotesize]{caption}
\usepackage[colorlinks=true, linkcolor=ikb, urlcolor=ikb, citecolor=ikb,
            anchorcolor=ikb]{hyperref}

\newcommand{\cP}{\ensuremath{\mathcal{P}}}
\newcommand{\cT}{\ensuremath{\mathcal{T}}}
\newcommand{\cPT}{\ensuremath{\mathcal{PT}}}

\begin{document}

\title{Exceptional Points and Stability in Nonlinear Models of Population Dynamics having $\mathcal{PT}$ symmetry
}
\author{Alexander Felski$^a$}
\email{alexander.felski@mpl.mpg.de}
\author{Flore K. Kunst$^{a,b}$}
\email{flore.kunst@mpl.mpg.de}
\affiliation{
$^a$ Max Planck Institute for the Science of Light, 91058 Erlangen, Germany\\
$^b$ Department of Physics, Friedrich-Alexander-Universit\"at Erlangen-N\"urnberg, 91058 Erlangen, Germany}

\begin{abstract}

Nonlinearity and non-Hermiticity, for example due to environmental gain-loss processes, 
are a common occurrence throughout numerous areas of science and lie at the root of many remarkable phenomena. For the latter, parity-time-reflection ($\mathcal{PT}$) symmetry has played an eminent role in understanding exceptional-point structures and phase transitions in these systems. Yet their interplay has remained by-and-large unexplored. 
We analyze models governed by the replicator equation of evolutionary game theory and related Lotka-Volterra systems of population dynamics. These are foundational nonlinear models that find widespread application, and offer a broad platform for non-Hermitian theory beyond physics.
In this context we study the emergence of exceptional points in two cases: (a) when the governing symmetry properties are tied to global properties of the models, and,  in contrast, (b) when these symmetries emerge locally around stationary states--in which case the connection between the local linear non-Hermitian model and an underlying global nonlinear system becomes tenuous.  
We outline further that when the relevant symmetries are related to global properties, the location of exceptional points in the linearization around coexistence equilibria coincides with abrupt global changes in the stability of the nonlinear dynamics. Exceptional points may thus offer a new local characteristic for the understanding of these systems.
Tri-trophic models of population ecology serve as test cases for higher-dimensional systems.

\end{abstract}
\maketitle

\section{Introduction}

Nonlinear dynamical systems are a widespread, if not ubiquitous, occurrence throughout numerous branches of natural science, with examples ranging from optical and atomic physics \cite{B2008,BG2015}, hydro- and thermodynamics \cite{K1975,C2020}, to chemical and biological processes \cite{S2000}, population ecology \cite{CL2020,SL1996,ML1975,KRF2002}, network science \cite{N1989,WBS2015,SF2006}, and socioeconomic systems \cite{BMS1997,HMH2002,T1993,G1982}.
Often, already relatively simple nonlinear systems act as cornerstone models that are capable of capturing central  qualitative behaviors and features, and give rise to varied phenomenology beyond what can be captured in the linear regime \cite{TS2002}.    

In comparison, non-Hermitian -- or pseudo-Hermitian -- theories have seen an extensive growth of interest rather recently. 
With modern developments commonly traced to the formative work on quantum mechanics with parity-time-reflection ($\cPT$) symmetry by Bender and B\"ottcher \cite{BB1998}, these systems 
have found particularly fruitful ground in the application to open system dynamics \cite{R2009,RPC2022}, especially in conjunction with topological features \cite{BBK2021, KEB2018,DFM2022,OS2022}.
Here, treating losses due to environmental interactions not as detrimental impediments that are to be minimized and avoided, but instead as advantageous features, which, together with controlled gain, can be beneficial and give rise to novel behaviors, has led to extensive theoretical studies and experimental implementations in various platforms \cite{CY2018,CS2024}, including mechanical, atomic and photonic systems.  

Parametric regimes in which nonlinear effects become significant are frequently present in these experiments \cite{GE2016,RKE2010,MME2008,MCS2018,MS2018,LPR2013,SLC2020,E2022,Y2021,MS2021,XKS2021,WRM2015,HDL2014,GKN2010}. They outline one natural path to investigations of the interplay between nonlinear dynamics and the features of non-Hermitian systems.
Another avenue emerges from applications of $\cPT$ symmetry and non-Hermitian topological effects in biological and chemical systems \cite{TAG2021,NT2024,ZT2024, BGG2016}.
Here, oscillatory dynamics are of particular interest, due to their prevalence in natural processes, and 
stability is of great importance for the self-sustained evolution of many-component systems.
Symmetry-based considerations and modelling have long played an important role in this field to determine characteristic properties, such as stability classifications. For instance, already the foundational Lotka-Volterra model of population dynamics exhibits an antisymmetric relation in its predator-prey interaction, that is between the gain of the predator and the loss of the prey population. Considering extensions of such systems, while retaining underlying symmetry properties, thus naturally matches the ideas lying at the heart of $\cPT$ symmetry.
In this context, examining the behavior of nonlinear systems through linearized dynamics in the vicinity of stationary states has been connected to the Schr\"odinger equation of quantum mechanics as a way of accessing results from non-Hermitian physics. The emergence of so-called exceptional points (EPs)--uniquely non-Hermitian singularities--and related non-Hermitian topological properties, such as the non-Hermitian skin effect \cite{BBK2021}, have recently been demonstrated in models of evolutionary game theory \cite{YMH2021,YMH2022,KGF2020,LDL2024,ZC2023}. 
This offers a new perspective on natural processes in biological, social, and economic systems through the lens of non-Hermiticity and topology.  

However, the linearization approach to stability analysis is ultimately limited to addressing the behavior of the stationary state. While resulting features commonly remain valid in small surrounding domains, the information gained is in general not applicable globally. 
In this paper, we compare and contrast the local and global properties of nonlinear $\cPT$-symmetric systems. In particular, we examine the connection of EPs in local linearizations with the behavior of the underlying nonlinear dynamics for extended evolutionary games of rock-paper-scissors, governed by the replicator equation, and for the topologically related generalized Lotka-Volterra systems originating in population ecology. 

Our discussion connects the following mechanisms. When the locally linearized model inherits a global symmetry of the nonlinear system, EP-type phase transitions are found to be indicative of global changes in the nonlinear dynamics, such as an abrupt global destabilization evolving toward the extinction of species or constituents. In contrast, we emphasize that the symmetries central to these non-Hermitian phase transitions, marked by EPs, can instead emerge as a local feature, disconnected from global symmetries of the full nonlinear theory. In this case, the position of the EP in the linearization decorrelates from the stability transitions of the nonlinear dynamics.

This study is structured as follows. 
Sections \ref{s2} and \ref{s3} briefly introduce and review the emergence of EPs in the context of evolutionary game theory through the example of the rock-paper-scissors game and $\cPT$-symmetric deformations thereof. The stability of the nonlinear dynamics is presented in comparison. By breaking the global symmetry of the model explicitly, differences between local and global symmetries are highlighted. 
In Secs.~\ref{s4} and \ref{s5} the connection to generalized Lotka-Volterra systems is used to analyze the changes in the global stability through Lyapunov functions, and in \ref{s6} these behaviors are examined for tri-trophic (that is three-dimensional) iterations of the system. We conclude in Sec.~\ref{s7}.

\section{Exceptional Points in Rock-Paper-Scissors Cycles}
\label{s2}

The game of rock-paper-scissors (RPS) is one of the quintessential mathematical models of game theory 
\cite{HS1998}. 
Players choose - rock, paper, or scissors - and compete pairwise: the outcomes are gathered in the so-called payoff matrix of the form
\begin{equation}
\label{eq0}
A = \,\,\,
    \begin{blockarray}{c c c l}
    \text{R} & \text{P} & \text{S} & \\
    \begin{block}{(c c c)l}
    0 & -1 & 1 & \,\,\,\, \text{R}\\
    1 & 0 & -1 & \,\,\,\,\text{P}\\
    -1 & 1 & 0 & \,\,\,\,\text{S}\\
    \end{block}
    \end{blockarray} ,
    \vspace{-0.4cm}
\end{equation}
which represents the success of one player's pure strategy (in rows) over that of the other (in columns): $1$ represents a win, $0$ a draw, and $-1$ a loss. 

In the context of evolution, one imagines a large well-mixed number of players following an initially fixed strategy, which then compete pairwise. For simplicity, we may imagine that after every game the loosing player adopts the winning strategy
\footnote{In general, winning players spawn duplicate players following their strategy according to some growth rate, while loosing players are culled according to some decay rate.}.
The dynamics of such a game, describing the dissemination of the possible strategies throughout the population of players, are governed by the replicator equation \cite{TY1978,HSS1979}
\begin{equation}
\label{eq1}
    \partial_t\, x_i = x_i [ (A \mathbf{x})_i - \mathbf{x}^T A \mathbf{x}] \, , 
    \quad
    i \in \{R,P,S\} \, ,
\end{equation}
where $\mathbf{x}$ encodes the frequency of strategies within the total number of players $n$, i.e., 
$\mathbf{x} = ( n_\text{R}/n,\,  n_\text{P}/n,\,  n_\text{S}/n )$.
The standard RPS dynamics are an illustrative example of \emph{cyclic dominance} \cite{SMJ2014}, in which the available strategies take turns in being prevalent, leading to cyclicity. At equidistribution of strategies, the dynamics are stationary, $\partial_t \mathbf{x}=0$, if not stably so.
In fact, the stability of this stationary \emph{coexistence point} $\Bar{\mathbf{x}}_\text{c} = (\tfrac{1}{3}, \tfrac{1}{3}, \tfrac{1}{3})$ is intuitively assessed by examining the dynamical reaction to an added small perturbation $\delta \mathbf{x}$. According to (\ref{eq1}), such a disturbance, the \emph{tangent flow} of the stationary state, evolves as 
\begin{equation}
\label{eq2}
    \partial_t \, \delta\mathbf{x} \approx \tfrac{1}{3} A\, \delta\mathbf{x} \,.
\end{equation}
The solution to this \emph{linearization} of the nonlinear replicator dynamics (\ref{eq1}) thus has the general form of a superposition of the three eigenstates $\mathbf{a}_j$ of the matrix $A$:
\begin{equation}
\label{eq3}
    \delta \mathbf{x}(t) = \sum_{j=1}^3 c_j \,\mathbf{a}_j\, \mathrm{e}^{\alpha_j t/3} \, ,
    \quad  \text{$c_j$ constant coefficients} \, .
\end{equation}
The growth or decay of small disturbances around $\Bar{\mathbf{x}}_c$ is therefore determined by the nature of the eigenvalues $\alpha_j$ of the payoff matrix - with only values having negative real part resulting in a refocusing to the coexistence point. For the standard game of RPS (\ref{eq0}), $\alpha_j \in \{\pm i \sqrt{3}, 0\}$, making its coexistence point a center of the linearized dynamics: disturbances neither grow nor decay, but persist. It is thus not asymptotically stable, but for its containment considered to be \emph{marginally stable}. 

This stability analysis of the evolutionary RPS game dynamics based upon a local linearization of the replicator dynamics exemplifies a general and ubiquitous approach to nonlinear models: \emph{Lyapunov's indirect method} \cite{K2013,W2010}. It is a special example in that the matrix $A$ governing the global dynamics of the system also determines the local linearized dynamics around the coexistence point. 
This is not generally so; it is, however, linked to the coexistence equilibrium lying at an equidistribution of strategies, which can always be achieved through a rescaling \cite{HS1998}. 
A major advantage of Lyapunov's indirect method is the structural simplicity of resulting in a linear first-order partial differential equation, which lends itself readily to analogies with the Schr\"odinger equation
\begin{equation}
\label{eq4}
    i \partial_t \ket{\psi} = H \ket{\psi}
\end{equation}
that governs the behavior of wave functions in quantum-mechanical systems.

In Ref.~\cite{YMH2021}, the authors draw exactly upon this resemblance to replicate novel behaviors found in the field of \emph{non-Hermitian} quantum physics within classical population dynamics.
Broadly speaking, one there extends models based on conventional \emph{Hermitian} descriptions, which guarantee entirely real eigenvalues that play the role of an energy spectrum of a persistent physical system, to \emph{pseudo-Hermitian} models, in which such real energies can also exist, but which furthermore allow for novel phase transitions into regimes with complex eigenvalues \cite{B2007, B2013}. 
The mathematical property of pseudo-Hermiticity is frequently associated with the presence of a physical symmetry within the underlying system, such as the seminal $\cPT$ symmetry: invariance under combined spacial reflection ($\cP$) and time reversal ($\cT$) \cite{BB1998, B2007}.

\begin{figure}[t]
\centering
\includegraphics[width=\columnwidth]
{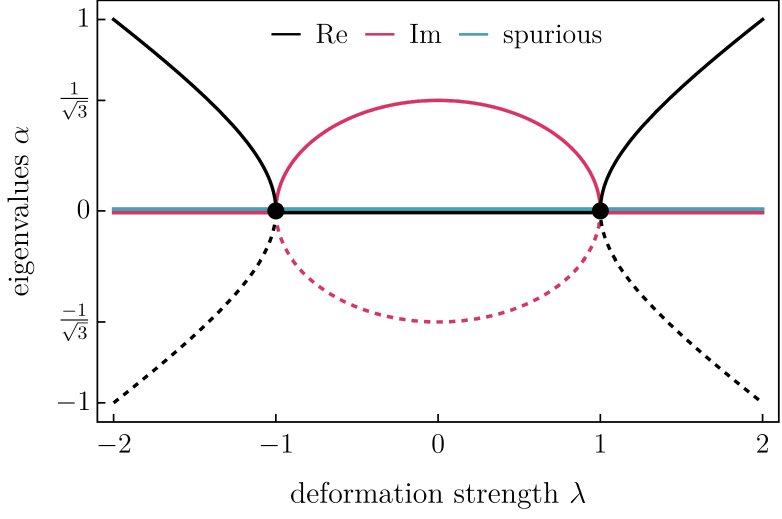}
\caption{
\label{f1}
Eigenvalues of the $\cPT$-symmetric payoff matrix $A_\lambda$ as a function of the deformation strength $\lambda$. EPs (black dots) arise at $\lambda_\text{EP}^\pm = \pm 1$, marking transitions in the nature of the coexistence equilibrium $\Bar{\mathbf{x}}_\text{c}$ from a center to a saddle point. 
}
\end{figure}

This is clearly illustrated by deforming the standard RPS game in the following way
\footnote{The model here differs slightly from \cite{YMH2022} for clarity, but follows their discussion in keeping with a coexistence point at equidistributed strategies.}:
\begin{equation}
\label{eq5}
A_\lambda = \,\,\,
    \begin{pmatrix}
        0 & -1 & 1 \\
        1& 0 & -1 \\
        -1 & 1 & 0
    \end{pmatrix}
    \, + \,
    \begin{pmatrix}
        0 & -\lambda & \lambda \\
        -\lambda& \lambda & 0 \\
        \lambda & 0 & -\lambda
    \end{pmatrix}
\, ,\quad \lambda \in \mathbb{R}
\, .
\end{equation}
For positive (negative) values of $\lambda$, the strategy $x_\text{R}$ wins more (less) against $x_\text{S}$, while $x_\text{S}$ loses less (more) against $x_\text{R}$. Similarly, $x_\text{R}$ loses more (less) against $x_\text{P}$, but $x_\text{P}$ wins less (more) against $x_\text{R}$. The outcomes of $x_\text{P}$ playing against $x_\text{S}$ remain unchanged, but in the case of a draw of $x_\text{P}$ both players obtain a payoff $\lambda$, while for a draw of $x_\text{S}$ both players obtain $-\lambda$. 
Both contributions to this payoff matrix uphold a symmetry of the linearized dynamics, determined by equation (\ref{eq2}) with $A_\lambda$, under combined time reversal $\cT$ ($t\rightarrow-t$) and parity reflection $\cP$, in the sense of an exchange of paper and scissors strategies; i.e., 
\begin{equation}
\label{eq6}
    \cP = \begin{pmatrix}
        1 & 0 & 0 \\
        0 & 0 & 1 \\
        0 & 1 & 0
    \end{pmatrix} 
    \, , \quad \text{s.t.} \quad
    \cP^2=\mathbbm{1}_3, \quad \cP A_\lambda \cP = -A_\lambda
    \, .
\end{equation}
Figure \ref{f1} shows the change within the eigenvalues of $A_\lambda$ as a function of the deformation strength $\lambda$ and indicates the nature of the equilibrium schematically: The initially imaginary values at $\lambda=0$ transition to appear in real-valued pairs with opposite sign (and an unchanged zero mode) after a critical strength $\vert\lambda_\text{EP}^\pm\vert=1$ is reached. The coexistence point destabilizes from a center at small $\lambda$ to a saddle point beyond the critical values.

This transition reflects a spontaneous breakdown of the underlying $\cPT$ symmetry in the linearized system: While the equation of motion (\ref{eq2}) is always symmetric, its solutions can lose this symmetry - and after this transition imaginary eigenvalues are no longer protected, see, e.g., \cite{B2007,B2013} for in-depth discussions.  
The critical coupling of this phase transition marks so-called \emph{exceptional points} having no counterpart in purely (anti-)Hermitian dynamics. They are deficient spectral degeneracies - meaning points of coalescing eigenvalues whose algebraic multiplicity, the number of coinciding eigenvalues, is larger than their geometric multiplicity, the number of linearly independent eigenvectors.
A topological $\mathbbm{Z}_2$-invariant,
\begin{equation}
\nu = \text{sgn}[\text{Disc}_\epsilon\, \text{det}(A_\lambda -\epsilon \mathbbm{1}) ]
= \text{sgn}[\lambda^2-1 ] \in \{\pm1\}
\, ,
\end{equation}
with $\text{Disc}_\epsilon$ referring to the discriminant of a polynomial in $\epsilon$,
can be attributed to the phases of spontaneously broken ($\nu=1$) and unbroken symmetry ($\nu=-1$) \cite{YMH2022}.

\begin{figure*}[t]
\centering
\includegraphics[width=0.9\textwidth]
{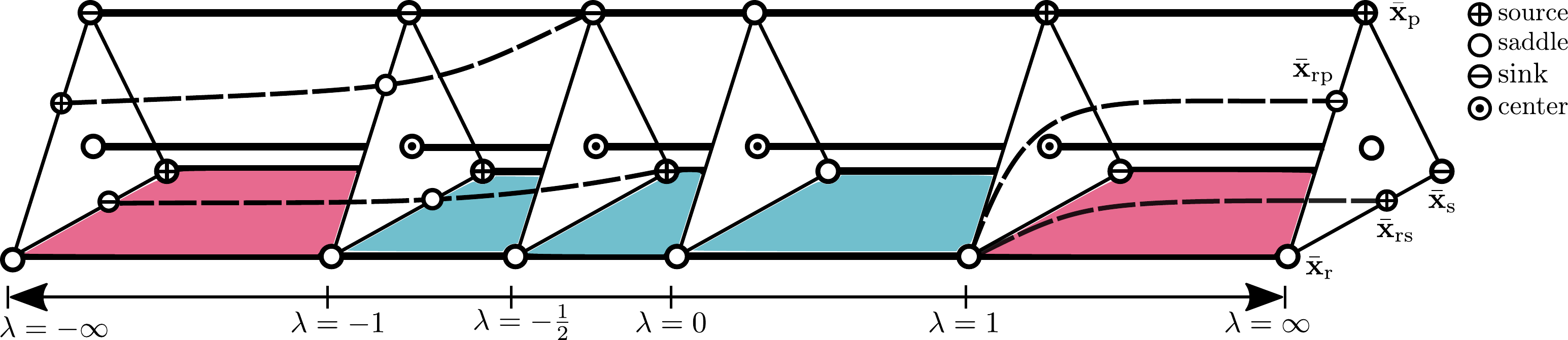}
\caption{
\label{f2}
Schematic visualization of the stationary states of the RPS dynamics and their nature on the 2-simplex as a function of the deformation strength $\lambda$. Shown are the globally unstable (red), and globally or regionally stable (blue) regimes, with exemplary dynamics in the top. The bifurcation of additional boundary equilibria is illustrated in dependence of $\lambda$ as dashed lines. 
}
\end{figure*}

We remark briefly on the minor difference between the linearized dynamics (\ref{eq2}) and the Schr\"odinger equation (\ref{eq4}) that is the appearance of the additional imaginary unit $i$: 
Due to this factor, stability of quantum systems governed by (\ref{eq4})
is commonly associated with \emph{real} energy eigenvalues of $H$, cf. (\ref{eq3}), while this finds its correspondence in \emph{imaginary} eigenvalues of $A_{\lambda}$ for (\ref{eq2}). As such, the simile of Hermitian quantum mechanics is an \emph{anti-}Hermitian payoff matrix.
However, extending this to the pseudo-Hermitian framework, neither matrix will be purely Hermitian nor purely anti-Hermitian - they are generally \emph{non-Hermitian}. And in both settings, we generally deform an initially stable system to become unstable after some underlying spontaneous symmetry breaking occurs.
The transfer of concepts from pseudo-Hermitian quantum physics to the study of nonlinear systems has been applied, e.g., in Refs.~\cite{GE2016,RKE2010,MME2008,MCS2018,MS2018,LPR2013,SLC2020,E2022,Y2021,MS2021}. In particular, the prospect of combining many such models within broken symmetry phases to realize the novel phenomenology of non-Hermitian topological states in biological systems has been explored \cite{TAG2021,NT2024,ZT2024}.
Nonetheless, these discussions ultimately build upon the linearized behavior within the stability analysis of their nonlinear constituents. 
However, this linearization is strictly valid only at the stationary position around which the system is perturbed. One may find extended validity within a local domain, that is in a regime of finite but likely only small perturbations - global properties, however, are all but ensured.

\section{Symmetries of local and global dynamics}
\label{s3}

While Lyapunov's indirect method provides a readily accessible technique to analyze the stability of equilibria in nonlinear systems, its meaningfulness relies on identifying the behavior of the nonlinear system with its linearization in the vicinity of a given stationary state. The applicability of this identification is captured in the Hartman-Grobman theorem \cite{W2010,HS1998,GH1983}, affirming it for \emph{hyperbolic} equilibria - that is, when no eigenvalue admits a vanishing real part. 
The coexistence point of the RPS dynamics given by (\ref{eq1}) and (\ref{eq2}), having a linearization with eigenvalues proportional to $\alpha_j \in \{\pm i \sqrt{3}, 0\}$, thus provides an illustrative example of a non-hyperbolic equilibrium. While we can classify this point as a center of its tangent flow, this does not ensure that it is also a center of the underlying nonlinear dynamics.
It might still attract or repel local trajectories sub-exponentially \cite{W2010}.
As such, more advanced techniques are required to assess even the behavior in close vicinity of such points, highlighting the precarious nature of transferring concepts from linear to nonlinear systems.  

That being said, the RPS coexistence point is in fact also a center of the replicator dynamics, and remains as such under the $\cPT$-symmetric deformation (\ref{eq5}) up to the critical strength, $\vert \lambda \vert < \vert \lambda_\text{EP} \vert = 1$, after which it turns into a saddle point and thus becomes unstable. 
We will return to the construction of Lyapunov functions that allow to demonstrate such marginal stability of the nonlinear dynamics throughout extended regions in Sec.~\ref{s5}.

With a connection between linearized and nonlinear dynamics in the vicinity of the coexistence point present, the non-Hermitian exceptional-point-type phase transition of the linearization also becomes an indicator of structural stability changes of the \emph{local} nonlinear dynamics.
Noting that the $\cPT$ symmetry, whose spontaneous breakdown underpins the exceptional point of the linearized system and its parallel to non-Hermitian quantum systems, is moreover a \emph{global symmetry} of the nonlinear system 
\begin{equation}
\label{eq7}
    \partial_t \, x_i = x_i [ (A_\lambda \mathbf{x})_i - \mathbf{x}^T A_\lambda \mathbf{x}] \, , 
    \quad \text{where} \,\,\,
    x_i = x_i(t)
    \, ,
\end{equation}
that is to say (\ref{eq7}) is obeyed also by
\begin{equation}
\label{eq8}
    \mathbf{x}' = \cPT\mathbf{x} = \big( x_\text{R}(-t), x_\text{S}(-t), x_\text{P}(-t) \big)^T 
    \, ,
\end{equation}
we may ask whether the exceptional points in the linearized dynamics can serve as an identifier of global stability changes within the full nonlinear system.

To examine the global behavior of the nonlinear dynamics, we characterize the remaining stationary states of the system:
while the coexistence point is a unique equilibrium \cite{HS1998}, the replicator dynamics (\ref{eq7}) admit additional boundary equilibria for which at least one type of strategy is not present in the player population.
In the standard RPS case ($\lambda=0$) three such points exist, namely the points for which the full player population follows a single strategy:
\begin{equation}
\label{eq9}
    \Bar{\mathbf{x}}_\text{r} = (1,0,0)^T \, , \quad
    \Bar{\mathbf{x}}_\text{p} = (0,1,0)^T \, , \quad
    \Bar{\mathbf{x}}_\text{s} = (0,0,1)^T  .
\end{equation}
The fixed-point nature of these points is ensured by the replicator equation, independent of the payoff matrix, and thus they remain equilibria of the deformed RPS dynamics with $A_\lambda$ as well.
For $\lambda>1$ or $\lambda<-\frac{1}{2}$ additional stationary states emerge at 
\begin{equation}
\label{eq10}
    \Bar{\mathbf{x}}_\text{rp} = 
    \big(\mfrac{2\lambda\!+\!1}{3\lambda}, \mfrac{\lambda\!-\!1}{3\lambda}, 0 \big)^T  
    \,\,\,\text{and} \,\,\,\,
    \Bar{\mathbf{x}}_\text{rs} = 
    \big(\mfrac{2\lambda\!+\!1}{3\lambda}, 0, \mfrac{\lambda\!-\!1}{3\lambda} \big)^T 
    .
\end{equation}
In the case of $\lambda<-\frac{1}{2}$, these bifurcate from $\Bar{\mathbf{x}}_\text{p}$ and 
$\Bar{\mathbf{x}}_\text{s}$, respectively, while for $\lambda>1$ they both (pitchfork-)bifurcate from $\Bar{\mathbf{x}}_\text{r}$, see Fig.~\ref{f2}. 

\begin{figure}[t]
\centering
\subfloat[\centering $\lambda = -1.5$]{
\includegraphics[width=0.46\columnwidth]
{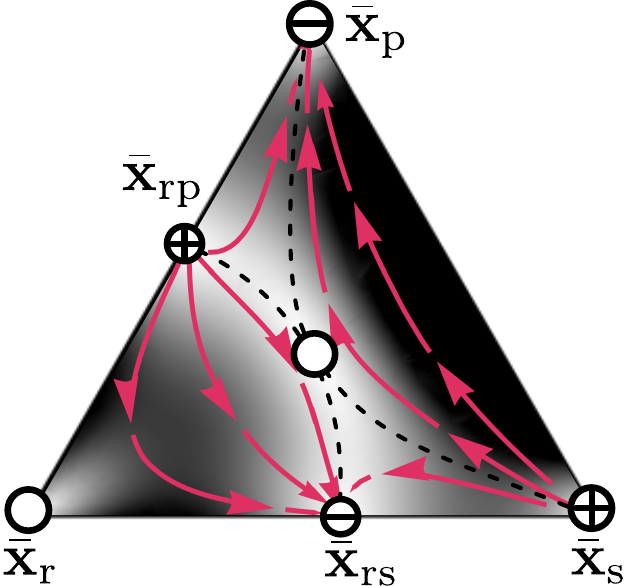}
\label{f2-2a}
}\hfill
\subfloat[\centering $\lambda = -0.75$]{
\includegraphics[width=0.46\columnwidth]
{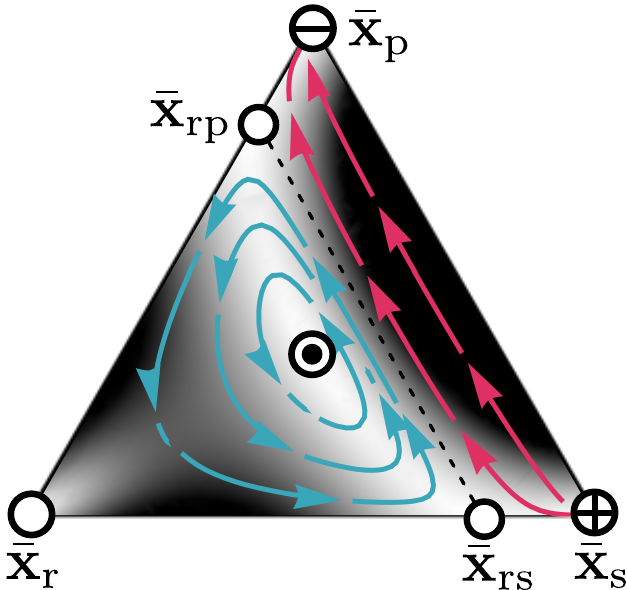}
\label{f2-2b}
}\\
\subfloat[\centering $\lambda = 0$]{
\includegraphics[width=0.46\columnwidth]
{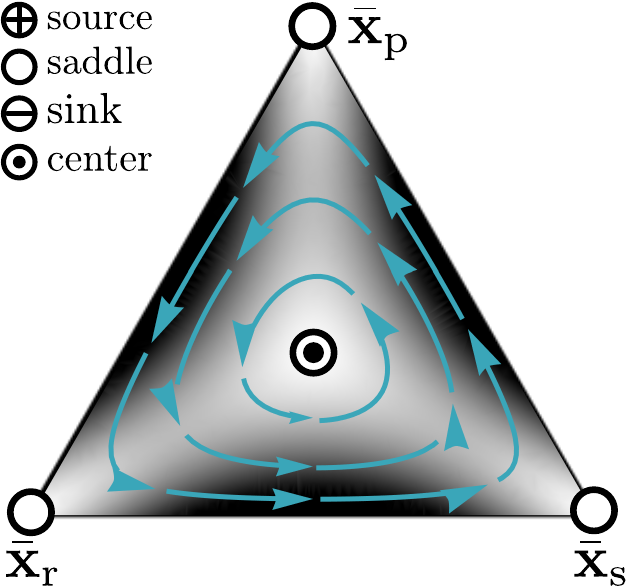}
\label{f2-2c}
}
\hfill
\subfloat[\centering $\lambda = 1.5$]{
\includegraphics[width=0.46\columnwidth]
{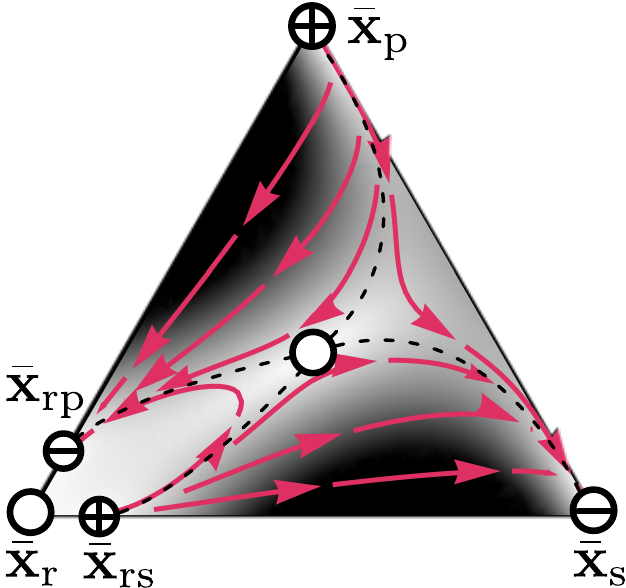}
\label{f2-2d}
}
\caption{
\label{f2-2}
RPS dynamics on the 2-simplex at exemplary values of the deformation strength $\lambda$. In (a) and (d) the dynamics are globally unstable. In (b) a regionally stable environment exist around the coexistence point. In (c) the dynamics are globally stable. High (low) velocities are indicated in white (black). 
}
\end{figure}

All boundary equilibria are hyperbolic stationary states and straightforwardly classified using Lyapunov's indirect method to be saddle points, sources, or sinks.
Figure~\ref{f2-2} shows the behavior of the $\cPT$-deformed RPS dynamics on the 2-simplex at representative deformation strengths $\lambda$ with the single-strategy equilibria as vertices. Indicated are in particular the position and nature of all equilibrium points.
Notably, the behavior is \emph{structurally stable} throughout different regimes of deformation, meaning changes in the parameter $\lambda$ do not change the type and stability of the critical points, and trajectories thus remain qualitatively unaffected \cite{W2010}.

In the standard RPS case and for all deformations within $-\frac{1}{2}<\lambda<\lambda_\text{EP}^+=1$, the boundary equilibria are saddles and the overall behavior follows cyclic dynamics around the center point at the coexistence equilibrium. These dynamics are stable globally (see also Sec.~\ref{s5}), cf. 
Fig.~\ref{f2-2c}. 

For $\lambda < -\frac{1}{2}$, the boundary equilibrium $\Bar{\mathbf{x}}_\text{r}$ remains a saddle, while $\Bar{\mathbf{x}}_\text{p}$ and $\Bar{\mathbf{x}}_\text{s}$ become a sink and a source, respectively. 
Accordingly, the dynamics can no longer be stable globally. 
However, both $\Bar{\mathbf{x}}_\text{p}$ and $\Bar{\mathbf{x}}_\text{s}$ furthermore bifurcate and give rise to additional boundary equilibria
$\Bar{\mathbf{x}}_\text{rp}$ and  $\Bar{\mathbf{x}}_\text{rs}$, which are saddle points within the regime $-1=\lambda_\text{EP}^-<\lambda < -\frac{1}{2}$. 
Here they are connected directly by a separatrix, indicated as dashed line in Fig.~\ref{f2-2b}, that separates unstable source-sink dynamics at small frequencies $x_\text{R}$ of the rock strategy within the player population from stable cyclic dynamics around the center coexistence point. As such, at $\lambda=-\frac{1}{2}$ the global stability of the dynamics continuously transitions to a regime in which extended regional stability of the system is maintained. 

Crossing the critical negative value $\lambda_\text{EP}^- = -1$, 
the boundary equilibria $\Bar{\mathbf{x}}_\text{rp}$ and  
$\Bar{\mathbf{x}}_\text{rs}$ on the 2-simplex edges change from being saddle points to a source and sink, respectively. 
The coexistence point destabilizes from a center to a saddle--its stable and unstable manifold, comprising the paths of steepest ascent and descent, form separatrices of the behavior of the system, indicated as dashed lines in Fig.~\ref{f2-2a}. 
These separatrices divide the dynamics into four regions, which are characterized by the combinations of sources and sinks that act as beginning and endpoints of all trajectories therein.
The dynamics \emph{abruptly} change from extended regional stability to become globally unstable in the regime 
$\lambda<\lambda_\text{EP}^- = -1$: all trajectories evolve toward one of the two sinks on the boundary.

\begin{figure}[t]
\centering
\subfloat{
\includegraphics[width=\columnwidth]
{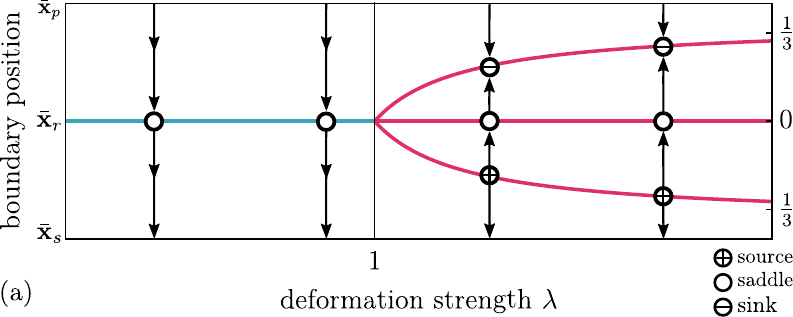}
\label{f3a}
}
\\[0pt]
\subfloat{
\includegraphics[width=\columnwidth]
{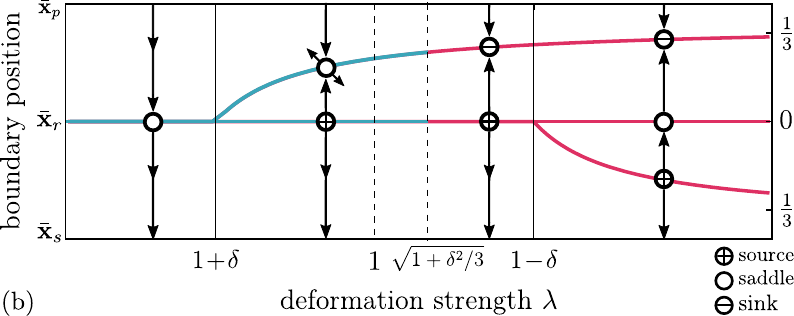}
\label{f3b}
}
\caption{
\label{f3}
Pitchfork bifurcation of the fixed boundary equilibrium $\Bar{\mathbf{x}}_\text{r}$ (a) without and (b) with defect. 
The globally unstable regime of deformation strengths $\lambda$ is indicated in red, while the globally or regionally stable regime is indicated in blue. 
}
\end{figure}

Similarly, beyond the positive critical value, $\lambda > \lambda_\text{EP}^+ = 1$, the boundary equilibria $\Bar{\mathbf{x}}_\text{p}$ and $\Bar{\mathbf{x}}_\text{s}$ turn from saddles to become a source and a sink, respectively--opposite to their behavior at large negative $\lambda$ values. 
$\Bar{\mathbf{x}}_\text{r}$ remains a saddle point, but with reversed ascent and descent paths, and pitchfork-bifurcates to give rise to the additional sink  $\Bar{\mathbf{x}}_\text{rp}$ and source $\Bar{\mathbf{x}}_\text{rs}$ on the 2-simplex edges, cf. Figs.~\ref{f2-2d} and \ref{f3a}. 
The coexistence point destabilizes from a center to a saddle and its stable and unstable manifold again form separatrices of the behavior of the system, indicated as dashed lines in Fig.~\ref{f2-2d}. The dynamics \emph{abruptly} become globally unstable in the regime 
$\lambda>\lambda_\text{EP}^+ = 1$.

Overall, one observes that the exceptional-point transitions at $\lambda_\text{EP}^\pm = \pm 1$ of the linearized dynamics at the coexistence point coincide with abrupt changes in the global stability of the nonlinear system. The regime of spontaneously-broken $\cPT$ symmetry of the linearization corresponds to a regime of globally unstable dynamics.
In the unbroken $\cPT$ symmetry regime of the linearization
an additional change in the structural stability of the global dynamics occurs at $\lambda=-\frac{1}{2}$; this change from global stability to extended regional stability, however, is continuous in the sense that the stable region smoothly expands to encompass the full global dynamics at this transition point. 

\subsection*{Explicitly broken symmetry}

We emphasize that the correlation of local EP-type phase transitions with global stability properties of the nonlinear dynamics is inherently tied to the $\cPT$ reflection being not only a symmetry of the linearized coexistence-point dynamics, but a \emph{global} symmetry of the nonlinear system.
This is elucidated by introducing an additional defect into the payoff matrix, which breaks this symmetry explicitly:
\begin{equation}
\label{eq11}
    A_\lambda^\delta = A_\lambda   +
    \delta \mathbbm{1}_3 \, ,
    \quad \delta \, \in \mathbb{R}
    \, ,
\end{equation}
due to the fact, that
\begin{equation}
\label{eq12}
   \cP \, \delta \mathbbm{1}_3 \, \cP =  \delta \mathbbm{1}_3 
   \,\,\, \rightarrow \,\,\,
   \cP  A_\lambda^\delta \cP \neq - A_\lambda^\delta
    \, ,
\end{equation}
in contrast to (\ref{eq6}).
Notably, this defect also preserves the position of the coexistence point at $\Bar{\mathbf{x}}_\text{c}$, and the boundary equilibria (\ref{eq9}) at $\Bar{\mathbf{x}}_\text{r}$, $\Bar{\mathbf{x}}_\text{p}$ and $\Bar{\mathbf{x}}_\text{s}$ persist due to the structure of the replicator equation. This makes it an ideal candidate to illustrate the impact of the explicit $\cPT$-symmetry breaking on the structural stability of the behavior, while preserving much of the underlying form of the system.
In an analogy to the Schr\"odinger equation picture, where an equivalent defect is purely imaginary, we will assume that $\delta<0$, corresponding to an overall loss, being commonplace for example in imperfect electromagnetic cavities.  
Such an instance has recently been investigated experimentally in a Kerr ring resonator \cite{HGF2024}.

Since the boundary equilibria remain hyperbolic stationary states, their classification proceeds as before, identifying them as saddle points, sources, or sinks.
The symmetry-breaking defect $\delta$ notably impacts the transitions between regimes of structural stability of the dynamics, with its most prominent impact on the pitchfork-bifurcation of $\Bar{\mathbf{x}}_\text{r}$ at $\lambda=1$.
In the presence of a defect $\delta$, this pitchfork splits into two staggered bifurcations at $\lambda=1\pm\delta$, see Fig.~\ref{f3}. 
This is to be expected, since pitchfork bifurcations are generically linked to the presence of an underlying symmetry, which disallows splittings such as that in Fig.~\ref{f3b} \cite{S2000}. 
Without defect, $\Bar{\mathbf{x}}_\text{r}$ is a fixed equilibrium, that is it maps to itself under the given symmetry ($\cPT$), which results in the pitchfork-bifurcating behavior. However, when this symmetry is explicitly broken, this behavior is no longer protected and breaks apart into two staggered (asymmetric) bifurcations.
Notably, $\Bar{\mathbf{x}}_\text{r}$ changes from a saddle point to a source in between the bifurcations. Moreover, the additional stationary state $\Bar{\mathbf{x}}_\text{rs}$, which bifurcates from $\Bar{\mathbf{x}}_\text{r}$ at $\lambda = 1+ \delta  <1$, is initially a saddle point, but transforms into a sink on the boundary at $\lambda=\sqrt{1+\delta^2/3}$. From there on, the behavior of the system becomes globally unstable.

Similarly, the bifurcations of $\Bar{\mathbf{x}}_\text{p}$ and $\Bar{\mathbf{x}}_\text{s}$, which are not fixed equilibria but map into one another under the $\cPT$ transformation, desynchronize from $\lambda=-\frac{1}{2}$ to $\lambda = -(1+\delta)/2$ and  $-(1-\delta)/2$, respectively. 
They do, however, not change their nature otherwise and remain transitions from saddle points to a source $\Bar{\mathbf{x}}_\text{p}$ at $\lambda=-(1+\delta)/2>-1/2$ and a sink $\Bar{\mathbf{x}}_\text{s}$ at $\lambda=-(1-\delta)/2<-\frac{1}{2}$. The latter destabilizes the systems' behavior from global stability to only regional stability, with an unstable regime at small frequencies of the strategy $x_\text{R}$. The additional bifurcating stationary states $\Bar{\mathbf{x}}_\text{rp}$ and $\Bar{\mathbf{x}}_\text{rs}$ are both initially saddle points, but become a sink and a source, respectively, at $\lambda=-\sqrt{1+\delta^2/3}$, which destabilizes the dynamics globally.

\begin{figure}[t]
\centering
\includegraphics[width=\columnwidth]
{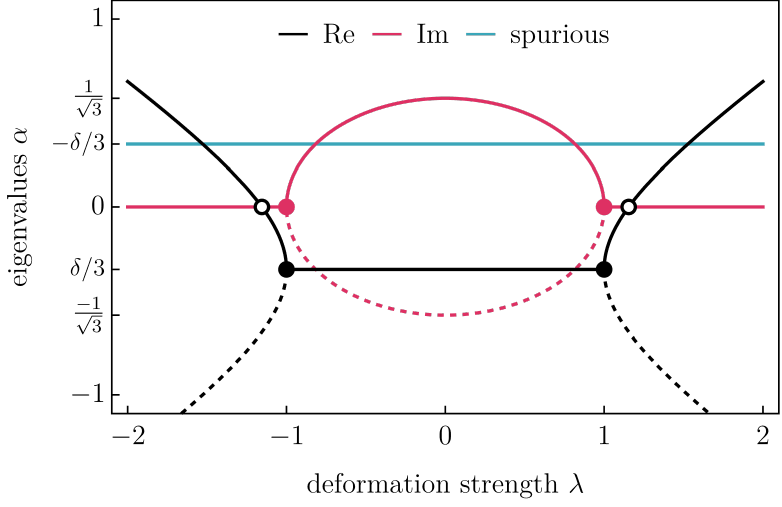}
\caption{
\label{f4}
Eigenvalues of the payoff matrix $A_\lambda^\delta$ as a function of the deformation strength $\lambda$. EPs arise at $\lambda_\text{EP}^\pm = \pm 1$ (black and red dots) as a result of passive $\cPT$-symmetry breaking.
The stability transition (black circles) of the deformed model does not coincide with the EPs due to the broken global symmetry.
}
\end{figure}

Overall, the addition of a small symmetry-breaking defect $\delta$ preserves qualitatively similar 
dynamics of the system: at large deformations $\vert\lambda\vert>>1$ the behavior becomes globally unstable, while it remains (at least regionally) stable at small deformation values. 
However, this behavior is most notably no longer correlated with the presence of an exceptional point in the tangent flow of the coexistence point $\Bar{\mathbf{x}}_\text{c}$, compare Fig.~\ref{f4}: global stability breaks down at $\lambda=\pm \sqrt{1+\delta^2/3}$ (black circles), which is a remnant of the approximate - but explicitly broken - global $\cPT$ symmetry in the model (\ref{eq11}). It is not associated with the exceptional point at $\lambda_\text{EP}=\pm 1$ (black and red dots) in the spectrum of the linearization at $\Bar{\mathbf{x}}_\text{c}$, proportional to the eigenvalues of $A_\lambda^\delta$
\begin{equation}
\label{eq13}
    \alpha_j \in \{ -\delta/3,\, \delta/3 \pm \sqrt{\lambda^2-1} \}
    \, .
\end{equation}

But if the $\cPT$ symmetry is broken, how can there still be an EP that reflects a spontaneous symmetry breaking transition at $\lambda_\text{EP}=\pm 1$?
This behavior is not tied to the global $\cPT$ symmetry with parity reflection (\ref{eq6}), but is instead rooted in a local property of the linearized dynamics at the coexistence point:
Due to the linearity of the governing equation (\ref{eq2}), the effect of any constant diagonal contribution to the governing matrix, such as the defect $\delta \mathbbm{1}_3$, factorizes in the resulting time evolution. Thus, the dynamics are comprised of an overall exponential growth or decay stemming from this diagonal matrix contribution, and a remainder which is governed by a $\cPT$-symmetric matrix. The latter thus imparts all the familiar spontaneous symmetry breaking behavior, including the presence of EPs. This property is sometimes referred to as \emph{passive $\cPT$ symmetry}, 
because it allows to probe $\cPT$ dynamics in lossy systems without active gain \cite{FXF2012}.
It is inherently tied to the linearity of the governing equation and the only $\cPT$-symmetry-breaking contribution in the governing matrix being a constant diagonal contribution.

We emphasize that the appearance of EPs in stationary-state dynamics of nonlinear systems can in general be tied to local or global symmetry properties 
\footnote{
Note that this differs from the studies \cite{HDL2014, GKN2010} of local and global stability in $\cPT$-symmetric nonlinear systems, wherein the behavior of the position of stationary states under $\cPT$ transformation is considered in a globally symmetric model.
In contrast, the spontaneous symmetry breaking transition, marked by an EP, arises in the local stability characterization of an isolated fixed-position equilibrium in this study. Moreover, the connection of global and local stability is analyzed for globally symmetric systems, as well as for systems in which such a symmetry is absent
}.
When they arise as manifestations of locally inherited global symmetries, we observe them to coincide with abrupt global changes in the stability of the nonlinear dynamics of the replicator equation (\ref{eq1}). However, when they arise due to local properties only, this correspondence no longer holds.
This highlights the precarious role of exceptional points when considering nonlinear systems.

\subsection*{Dimensionality}

Lastly, we comment on the dimensionality of the rock-paper-scissors dynamics described. While players choose between three different strategies, the degrees of freedom are restricted by the fact that the replicator equation governs the frequencies of strategies within the player population. As such, the dimension of the model is geometrically constrained by the condition $x_\text{R}+x_\text{P}+x_\text{S}=1$, which limits the dynamics to a two-dimensional domain: the 2-simplex, e.g., used for visualization in Fig.~\ref{f2}.
In the tangent flow around a stationary state, this geometric constraint is, however, not immediately apparent. The solution of the three-dimensional linear time-evolution equation needs to be restricted to the two contributions from eigenvectors within the plane of the 2-simplex. One eigenvalue of the linearization thus becomes spurious and does not contribute to the classification of the equilibrium points. 
In the tangent flow of the coexistence equilibrium, this eigenvalue corresponds to the flat-band solution, compare Figs.~\ref{f1} and \ref{f4}.
Notably, the EPs within these linearizations are not so-called \emph{third-order exceptional points} (EP3s), at which the geometric multiplicity is $1$ and three linearly independent eigenvectors reduce to a single eigenvector, despite the initial spectral similarity \cite{DJH2021,MB2021}. The spurious flat-band solution remains a completely sterile and linearly independent result, and the dynamics within the 2-simplex plane show a conventional second-order square-root EP.

Given that the evolutionary game theory dynamics of the rock-paper-scissors game are a two-dimensional model, one may ask to find a topologically equivalent two-variable system description. In fact, such a model is given by the 
well-established (generalized) Lotka-Volterra model of population dynamics \cite{HS1998}. We provide a brief overview of this correspondence in the following section and utilize it to examine the stability $\cPT$-symmetric deformations of the system through the construction of Lyapunov functions. Moreover, the framework of the generalized Lotka-Volterra model provides a natural way to extend the study of exceptional points in nonlinear systems to higher dimensions, while retaining a basis in the context of population ecology.

\section{Coexistence in the Lotka-Volterra Model}
\label{s4}

The generalized Lotka-Volterra (GLV) system provides a cornerstone model for the joint evolution of $N$ species populations $y_i\in \mathbbm{R}^{\geq0}$ \cite{HS1998}. 
Its dynamics are governed by a system of nonlinear first-order differential equations,
\begin{equation}
\label{eq15}
    \partial_t y_i(t) = r_i \,y_i(t) + y_i(t) \big[  \sum_{j=1}^N b_{ij}\, y_j(t) \big] ,
\end{equation}
where $r_i$ denote the intrinsic growth rates of the species and $b_{ij}$ specify their pairwise interactions. 
A matrix form is given by
\begin{equation}
\label{eq16}
    \partial_t \mathbf{y}(t) = \text{diag}(y_1(t),...,y_N(t))\, [ \mathbf{r} + B\, \mathbf{y}(t)] ,
\end{equation}
where $B=(b_{ij})$ is called the \emph{interaction matrix}.
The pairwise interactions of species may be competitive ($b_{ij} \,\&\, b_{ji} < 0$), mutualistic ($b_{ij} \,\&\, b_{ji} > 0$), or antagonistic ($b_{ij} b_{ji} < 0$, predator-prey type) and self-interactions can be logistic (self-regulating through negative density dependence) for $b_{ii}<0$, or orthologistic (positive density dependence) for $b_{ii}>0$.
Such systems admit a large variety of possible dynamical behaviors, including limit cycles, chaotic evolution and stationary fixed points.  
In particular the cyclic evolution possible in antagonistic models has long been of wide interest in the context of theoretical ecology due to its prevalence in population dynamics and chemical kinetics (self-catalyzing reactions), which provided the foundation for the original two-species model \cite{L1920,V1926}.
Today, the GLV system finds widespread application beyond these fields, ranging from neural nets \cite{N1989,Y2010} and plasma physics \cite{LP1975} to economic systems \cite{OS2021,G1982}.

A notable property of the system (\ref{eq15}) is the topological equivalence of its dynamics to those of the $(N+1)$-dimensional replicator equation \cite{HS1998}, cf.~(\ref{eq1}), with a geometric constrain in the sum of frequencies.
By supplementing the GLV system with an ancillary variable $y_{0}(t)=1$ and performing a barycentric coordinate transformation $\mathbf{y}(t)\rightarrow\mathbf{x}(t)$ so that:
\begin{equation}
\label{eq17}
   x_i(t) = \frac{y_i(t)}{\sum_{j=0}^{N} y_j}  \, ,
   \quad\text{inversely}\quad
   y_i(t) = \frac{x_i(t)}{x_{0}} \, ,
\end{equation}
one arrives at
\begin{align}
\label{eq18}
   \partial_t x_i =& x_i \big[\mfrac{\partial_t x_0}{x_0}
   + \sum_{j=0}^N \Tilde{B}_{ij} (\mfrac{x_j}{x_0}) \big]\\
   =& \mfrac{1}{x_0} \{ x_i \big[
   (\Tilde{B}\mathbf{x})_i - \mathbf{x}^T \Tilde{B}\mathbf{x}] \}
   \, ,
\end{align}
where the matrix $\Tilde{B}$ has the form
\begin{equation}
\label{eq20}
    \Tilde{B} = 
    \left( \begin{array}{@{}c|c@{}}
      \,0 & 
    \begin{matrix}
    0 & \cdots & 0 \,
   \end{matrix}\\ 
    \hline
    \,\mathbf{r} & 
    B
   \\
\end{array} \right) 
\, .
\end{equation}
This coincides with the replicator equation up to a scaling of the velocity, that is the trajectories are topologically equivalent to those of the replicator equation with payoff matrix $\Tilde{B}$, but they are traversed at different timescales \cite{HS1998}.
This embedding of the $N$-dimensional GLV dynamics in an $(N+1)$-dimensional replicator system establishes the correspondence of the models in the context of stability analysis.

One further notes that the structure of the payoff matrix in the replicator equation, cf. (\ref{eq1}), has the following ambiguity \cite{HSS1979}: The addition of a constant to any column of the payoff matrix,
\begin{equation}
\label{eq21}
    A \rightarrow A'=
    \begin{pmatrix}
    a_{00} + c_0 & a_{01} + c_1 & \cdots & a_{0N}+ c_N \\
    a_{10} + c_0 & a_{11} + c_1 & \cdots & a_{1N}+ c_N \\
    \vdots & \vdots &  & \vdots \\
    a_{N0} + c_0 & a_{N1} + c_1 & \cdots & a_{NN}+ c_N
   \end{pmatrix}\\ 
\, ,
\end{equation}
leaves the replicator dynamics unchanged. As such, any payoff matrix can always be brought into the form (\ref{eq20}) and an equivalent GLV system can thus be found. In this sense, the geometrically constrained replicator equation can be regarded as a compactification of the GLV system.

In consequence, the two-dimensional antagonistic GLV system of the form
\vspace{-0.2cm}
\begin{equation}
\label{eq22}
    \begin{alignedat}{2}
    & \partial_t y_1 = y_1 ( 1 + y_1 - 2 y_2 ) ,\\
    & \partial_t y_2 = y_2 (-1 +2y_1 - y_2 ) 
    \end{alignedat}
    \vspace{-0.1cm}
\end{equation}
provides a two-variable description that corresponds to the dynamics of the game of rock-paper-scissors (\ref{eq0}) on the 2-simplex. Notably, the predator species $y_2$ self-regulates due to a logistic self-interaction term and the evolution of the prey species $y_1$ contains an orthologistic (positive density dependence) self-interaction term.  
One again finds a unique coexistence equilibrium, sometimes called \emph{rest point}, at $\mathbf{y}=(1,1)$, around which the dynamics are governed by the linearization
\vspace{-0.2cm}
\begin{equation}
\label{eq23}
     \partial_t \, \delta\mathbf{y} \approx B\, \delta\mathbf{y} \quad \text{with} \quad
     B = \begin{pmatrix}
          1 & -2 \\
          2 & -1
     \end{pmatrix}
\, .
\vspace{-0.1cm}
\end{equation}
This behavior is entirely determined by the interaction parameters of the population dynamics model in the interaction matrix $B$ of the full nonlinear system.
Note that in this two-variable description, the spurious flat-band solution in the spectrum of the linearization is no longer present.

In addition to linking the evolutionary game theory results to models of population dynamics and thus providing a wider context for non-Hermitian physics in nonlinear systems, the topological equivalence between the replicator and GLV systems allows for the transfer of established approaches between these frameworks. 
In the following, we utilize this equivalence to determine Lyapunov functions that establish stability properties of the models considered. 
Specifically, we will reexamine the connection between EPs arising in the local dynamics at the coexistence point and global stability properties through this lens.

\section{Symmetry and Stability in the Two-Species Model}
\label{s5}

\begin{figure*}[t]
\centering
\subfloat[\centering $\lambda = -1.5$]{
\includegraphics[width=0.23\textwidth]
{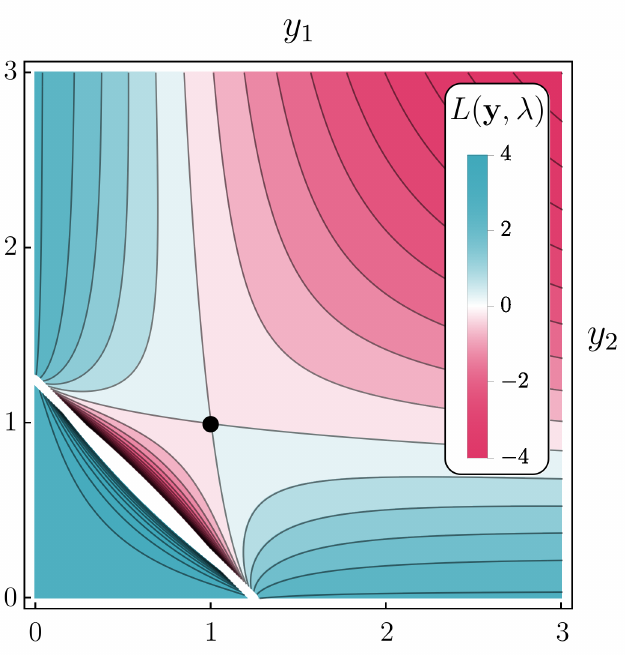}
\label{f5a}
}\hfill
\subfloat[\centering $\lambda = -0.75$]{
\includegraphics[width=0.23\textwidth]
{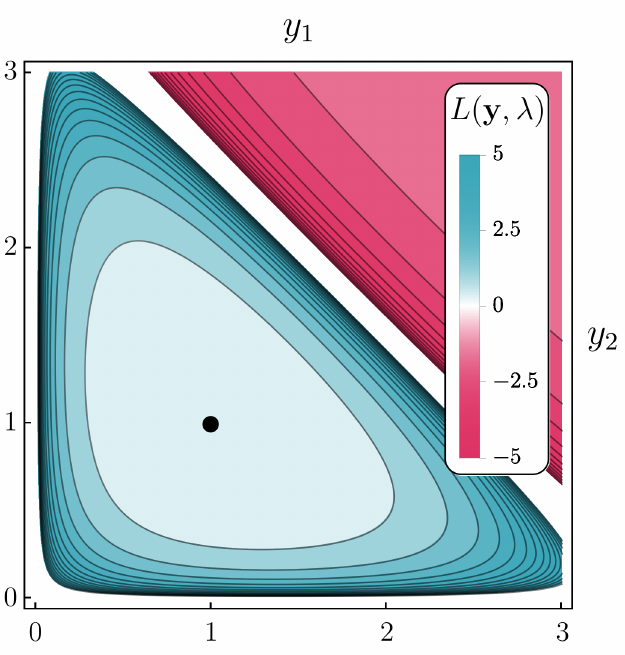}
\label{f5b}
}\hfill
\subfloat[\centering $\lambda = 0$]{
\includegraphics[width=0.23\textwidth]
{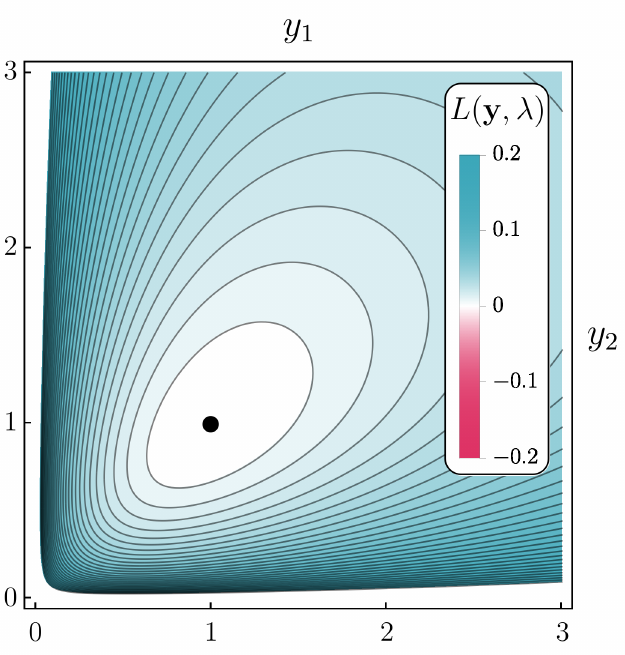}
\label{f5c}
}
\hfill
\subfloat[\centering $\lambda = 1.5$]{
\includegraphics[width=0.23\textwidth]
{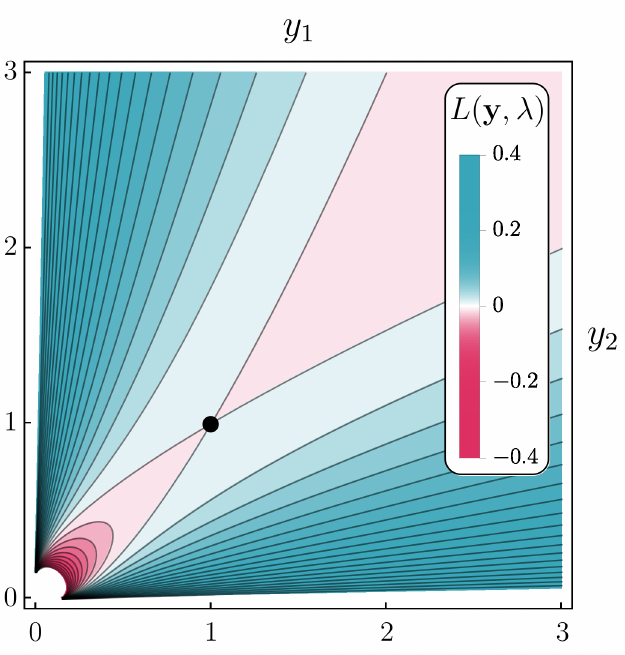}
\label{f5d}
}
\caption{
\label{f5}
Amplitude of the Lyapunov-candidate function $L(\mathbf{y},\lambda)$  for different values of the deformation strength $\lambda$.
The behavior in (a) is representative of $\lambda < -1$,
in (b) of $\lambda \in (-1,-\tfrac{1}{2})$, in (c) of $\lambda \in (-\tfrac{1}{2},1)$, and in (d) of $\lambda>1$. The coexistence point is shown as a black dot.
}
\end{figure*}

Determining the stability of an equilibrium point in a nonlinear system can, at times, be challenging. 
For non-hyperbolic stationary states, the equivalence of the tangent flow and the underlying nonlinear dynamics is not ensured and Lyapunov's indirect method of investigating the linearization around the equilibrium is unviable.
A more robust technique lies in \emph{Lyapunov's direct method} \cite{K2013,W2010}, 
which utilizes the analog of a potential function, a so-called Lyapunov function, to establish stability.
The existence of such a function $L: \mathbbm{R}^N \rightarrow \mathbbm{R}$, satisfying $L(\Bar{\mathbf{y}}) = 0$ and $L(\mathbf{y}) > 0$ if $\mathbf{y}\neq\Bar{\mathbf{y}}$, with $\tfrac{\mathrm{d}}{\mathrm{d}t} L(\mathbf{y}) \leq 0$ in a neighborhood of $\Bar{\mathbf{y}}$, ensures the marginal stability of the equilibrium point $\Bar{\mathbf{y}}$ \cite{W2010}. The difficulty then lies in determining this Lyapunov function, since a general construction remains unknown. 
However, for many specific cases examples have been established and the connection between the replicator equation and GLV system illustrates how such examples may be transferred in order to assert stability properties. 

Following the construction in Sec.~\ref{s4}, the globally $\cPT\!$-symmetric deformation (\ref{eq5}) of the rock-paper-scissors model of evolutionary game theory is topologically equivalent to the two-dimensional GLV system with growth rates and interaction matrix
\begin{equation}
\label{eq}
    \mathbf{r} = (1-\lambda, \, \lambda -1)^T 
    \quad \text{and} \quad
    B = \begin{pmatrix}
        1+2\lambda & -2-\lambda \\
        2+\lambda & -1-2\lambda
    \end{pmatrix}
\, ,
\end{equation}
that is
\begin{equation}
\label{eq25}
    \begin{alignedat}{2}
    & \partial_t y_1 = y_1 [ (1-\lambda) + (1+2\lambda) y_1 - (2+\lambda) y_2 ] ,\\
    & \partial_t y_2 = y_2 [(\lambda-1) +(2+\lambda) y_1 - (1+2\lambda) y_2 ] .
    \end{alignedat}
\end{equation}
Here the dependence on the strategy $x_\text{R}$ was eliminated as ancillary variable and the strategies $x_\text{P}$ and $x_\text{S}$ were transformed to $y_1$ and $y_2$, compare (\ref{eq17}). Accordingly, one again finds the resulting nonlinear system to be globally symmetric under combined time reversal, $\cT:t\rightarrow-t$, and 
parity reflection $\cP$, in the sense of an exchange $y_1 \leftrightarrow y_2$; that is 
\vspace{-0.2cm}
\begin{equation}
\label{eq26}
    \cP = \begin{pmatrix}
        0 & 1 \\
        1 & 0
    \end{pmatrix} 
    \, .
\end{equation}

A Lyapunov function for this model can be found as follows.
Starting from the replicator equation with payoff matrix (\ref{eq5}), we utilize the freedom to add constant column contributions, see (\ref{eq21}), without changing the dynamics, in order to eliminate the diagonal entries of the matrix:
\begin{equation}
\label{eq27}
    A_\lambda \rightarrow A_\lambda'=
    \begin{pmatrix}
    0 & -1-2\lambda & 1+2\lambda \\
    1-\lambda & 0 & \lambda-1 \\
    \lambda - 1 & 1-\lambda & 0
   \end{pmatrix}\\ 
\, .
\end{equation}
In this form the payoff matrix is antisymmetrizable: given the diagonal matrix 
$D = d\cdot  \text{diag}(\tfrac{1+2\lambda}{1-\lambda}, 1,1 )$ with $d \in \mathbbm{R}^{>}$, one finds that 
\vspace{-0.1cm}
\begin{equation}
\label{eq27}
   (A_\lambda' D) = - (A_\lambda' D)^ T
\, .
\vspace{-0.1cm}
\end{equation}
However, antisymmetrizability necessitates $D$ to be positive, which holds only for $\lambda \in [-\tfrac{1}{2}, 1]$.
In this regime, the replicator equation then admits a constant of motion of the form
\vspace{-0.2cm}
\begin{equation}
\label{eq29}
   C(\mathbf{x}, \lambda) = 
   \frac{
   \big[x_\text{R}\,  \big(\mfrac{1-\lambda}{1+2\lambda} \big) \big]^{\tfrac{1-\lambda}{3(1+\lambda)}} 
   \,\, \big(x_\text{P}\, x_\text{S} \big)^{\tfrac{1+2\lambda}{3(1+\lambda)}}
   }
    {x_\text{R}\,  (\tfrac{1-\lambda}{1+2\lambda}) + x_\text{P} + x_\text{S}}
\, ,
\end{equation}
compare \cite{K2004}. This can now be transformed into a constant of motion of the topologically equivalent GLV system (\ref{eq25}) using (\ref{eq17}), yielding 
\begin{equation}
\label{eq30}
   C(\mathbf{y}, \lambda) = 
   \frac{
   (y_1\, y_2 )^{\tfrac{1+2\lambda}{3(1+\lambda)}} 
      }
    { (\tfrac{1-\lambda}{1+2\lambda}) + y_1 + y_2}
    \,\, \big(\mfrac{1-\lambda} {1+2\lambda}\big)^{\tfrac{1-\lambda}{3(1+\lambda)}}
\, .
\end{equation}
In general, this establishes a (noncanonical) Hamiltonian structure \cite{W2010},
\vspace{-0.2cm}
\begin{equation}
\label{eq31}
   \Dot{\mathbf{y}} = J \,\nabla  C(\mathbf{y}, \lambda) \, , \quad
   \text{with} \quad 
   J = \begin{pmatrix}
       0 & f \\-f & 0
   \end{pmatrix},
\vspace{-0.2cm}
\end{equation}
where
\vspace{-0.2cm}
\begin{equation}
    f = \frac{
    (y_1\, y_2 )^{\tfrac{1+2\lambda}{3(1+\lambda)}} \,
    \big(\tfrac{1-\lambda} {1+2\lambda}\big)^{\tfrac{1-\lambda}{3(1+\lambda)}}
    }
    { 
    3(1+\lambda) \,y_1 y_2 \,
    \big[(\tfrac{1-\lambda}{1+2\lambda}) + y_1 + y_2\big]^2
    }
\end{equation}
giving rise to the equations of motion (\ref{eq25}).
However, this constant of motion has one essential downside: outside the range $\lambda \in [-\tfrac{1}{2},1]$ it becomes a complex-valued function. Since we are interested in constructing a Lyapunov-candidate function based on the constant of motion,
finding a real-valued expression is desirable. 
Nevertheless, this complex nature notably originates in the last factor of (\ref{eq30}), which arises as a contribution from the introduction of the ancillary variable in the compactification of the GLV equation to the replicator form, and which thus only contributes as an overall factor. As such, we may drop this contribution without affecting the time-independence of $C(\mathbf{y},\lambda)$. 
The resulting function is real-valued for all values of $\lambda$, and we can consider it as an energy function of the population dynamics
\vspace{-0.1cm}
\begin{equation}
    H(\mathbf{y}, \lambda) = \frac{
    (y_1\, y_2 )^{\tfrac{1+2\lambda}{3(1+\lambda)}} 
      }
    { (\tfrac{1-\lambda}{1+2\lambda}) + y_1 + y_2}
\, .
\vspace{-0.1cm}
\end{equation}
Based on this Hamiltonian $H$, we can now introduce a Lyapunov-candidate function as 
\begin{equation}
    L(\mathbf{y},\lambda) = H( \mathbf{1} ,\lambda)- H(\mathbf{y},\lambda).
\end{equation}
Since this function is derived from a constant of motion, $\frac{\mathrm{d}}{\mathrm{d}t}L(\mathbf{y},\lambda)=0$.
Figure~\ref{f5} shows the amplitude of $L(\mathbf{y},\lambda)$ for exemplary values of the deformation strength $\lambda$.
In the vanishing-deformation limit $\lambda=0$, see Fig.~\ref{f5c}, this function is globally strictly positive with  global minimum at the coexistence equilibrium. It thus establishes the \emph{global} (marginal) stability of the theory. In fact, this remains the case for any deformation strength with value $\lambda \in [-\tfrac{1}{2},1]$. 
For $\lambda \in (-1, -\tfrac{1}{2})$, the coexistence equilibrium becomes a local minimum of the function, see Fig.~\ref{f5b}. $L(\mathbf{y},\lambda)$ remains strictly positive in a finite region with the boundary given by the separatrix connecting the two additional boundary saddle points, compare Sec.~\ref{s3}. This establishes regional stability around the coexistence equilibrium with a breakdown at large population sizes. 
For deformation strengths $\vert \lambda \vert > 1 $, see Fig.~\ref{f5a} and Fig.~\ref{f5d}, the function is no longer meeting the requirements of a Lyapunov function, even within a small neighborhood of the coexistence equilibrium, indicating the destabilization of the theory - coinciding with the deformation strength crossing the EP within the linearization.  

Overall, the approach using Lyapunov's direct method confirms the presumed connection between the linearized and nonlinear dynamics in the vicinity of the coexistence point for both the rock-paper-scissors model and for the topologically equivalent two-dimensional GLV system.

\section{Tri-Trophic Food Webs}
\label{s6}

While the two-species GLV system captures various interaction types between two constituents, these dynamics are often modified by the presence of further species. This results in potentially extensive interaction networks or, in the context of ecology, food webs \cite{ABC2024}.
Despite the complexity of such networks in nature, their central dynamical features are commonly controlled by relatively few dominant connections \cite{BSH2017}.  
Grouping species categorically into trophic levels, indicating their location within a food chain, has also turned out to be a valuable concept that provides a comparability between vastly different ecosystems \cite{Y2001}.
Three-dimensional models form a fundamental component of these extensive ecosystems and can thus serve as instructive tool for building a mechanistic understanding of ecological dynamics \cite{APF2019}.
Moreover, many agricultural ecosystems comprise only simplified biodiversity for which tri-trophic models can be valuable tools with potential application, e.g., in improved pest control strategies \cite{LLA2009}.  
As such, three-species food webs provide essential models for a qualitative understanding of ecological dynamics. They can admit various structures, including (linear) food chains, chains with omnivory, or cyclic food chains.

In this section, we demonstrate the connection between EPs in the linearized dynamics around a coexistence point with changes in global stability for tri-trophic models as an example of its application to higher-dimensional GLV systems.
To this end, we introduce the following family of three-species GLV systems with growth rate and interaction matrix of the form:
\begin{figure}[t]
\centering
\subfloat[]{
\includegraphics[width=0.28\columnwidth]
{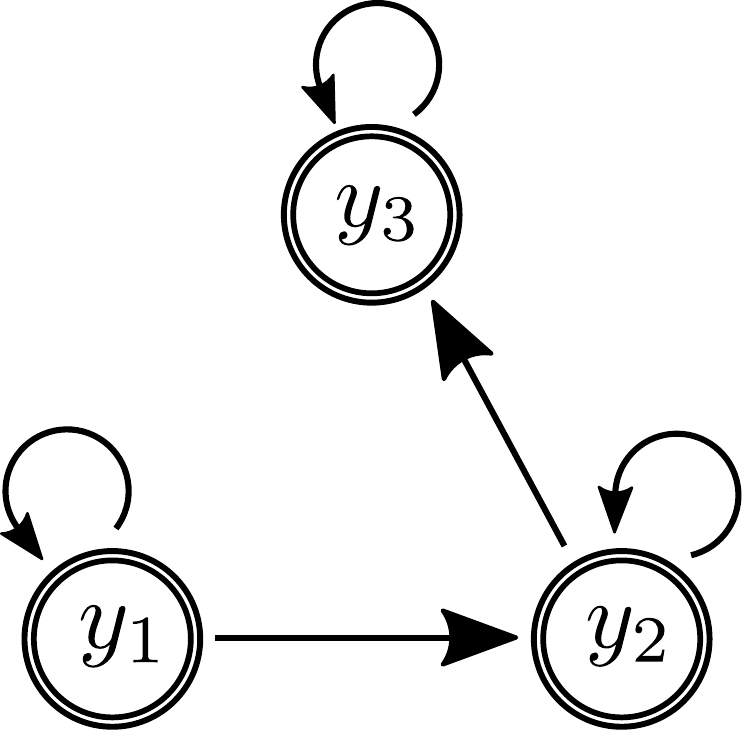
}
\label{f6a}
}
\hfill
\subfloat[]{
\includegraphics[width=0.28\columnwidth]
{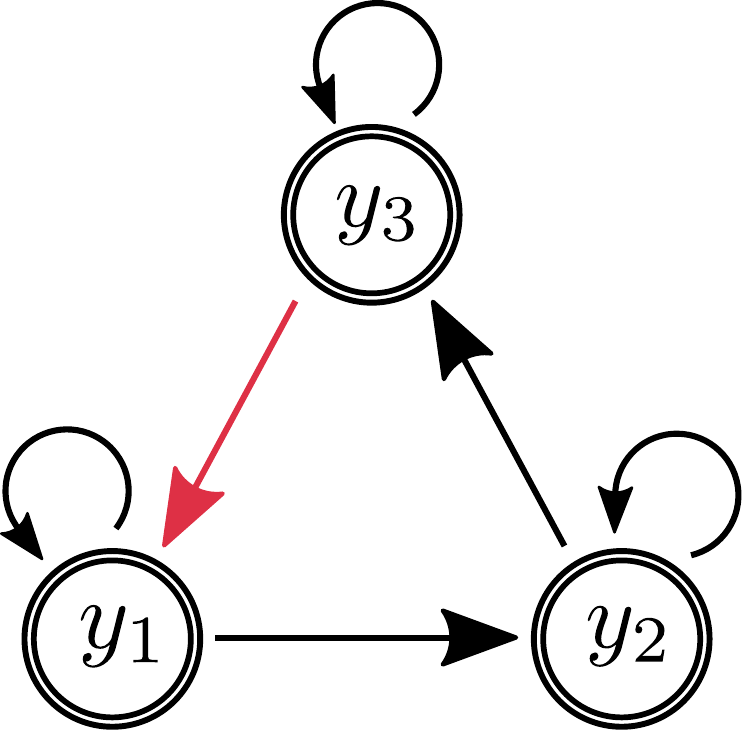
}
\label{f6b}
}
\hfill
\subfloat[]{
\includegraphics[width=0.28\columnwidth]
{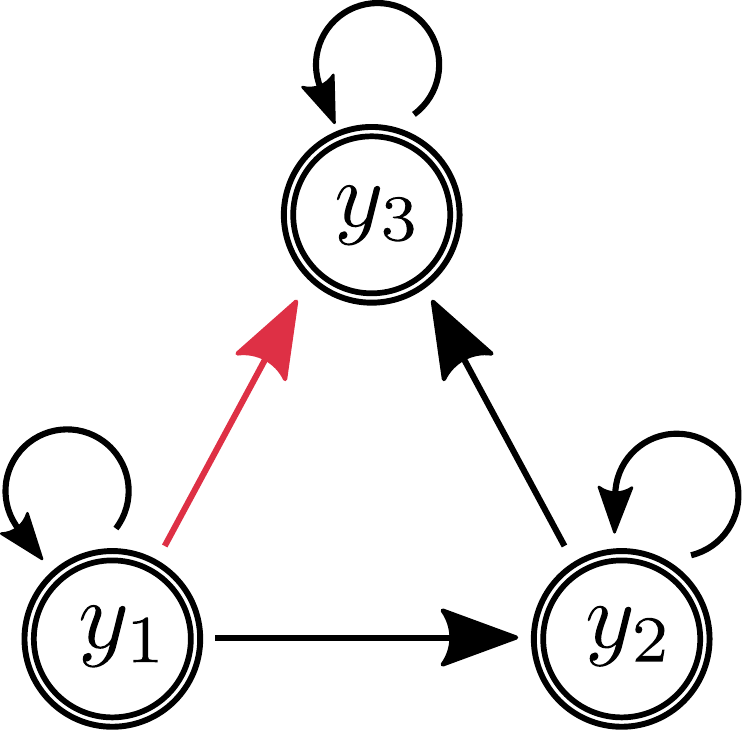
}
\label{f6c}}

\caption{
\label{f6}
(a) Food chain, (b) cyclic food chain, and (c) food chain with omnivory, distinguished by the $y_1$-$y_3$ interaction in red.
}
\end{figure}

\begin{widetext}
\begin{equation}
\label{eq34}
    \mathbf{r} = \frac{1}{\mathcal{N}} \,\,
    (1,0,-1)^T
    \quad \text{and} \quad
    B = \frac{1}{\mathcal{N}} \,\,
    \begin{pmatrix}
        1 & (-2\!-\!\delta) & \delta \\
        (3\!+\!\delta) & 0 & (-3\!-\!\delta) \\
        -\delta & (2\!+\!\delta) & -1 
    \end{pmatrix}
    + \frac{\lambda}{\sqrt{3}}
    \begin{pmatrix}
        1 & -1 & 0 \\
        -1 &  0 & 1 \\
        0 & 1 & -1 
    \end{pmatrix}
,
    \quad \text{where}\,\,\mathcal{N}=\sqrt{11+10\delta+3\delta^2}
\,.
\end{equation}
\end{widetext}

\noindent
Choosing the parameter $\delta =0$ results in a food chain model, while positive or negative values result in a cyclic food chain or introduce omnivory, respectively. We choose $\delta \in \{0, \pm 1\}$ in the following to illustrate the behavior in these different types of systems. 
As in the two-dimensional toy model, the system admits a coexistence equilibrium, scaled to lie at $\Bar{\mathbf{y}}_\text{c} = (1,1,1)$. For simplicity, the normalization of the matrix contributions in (\ref{eq34}) are chosen such that the linearization around this equilibrium has eigenvalues $\beta_j \in \{0, \pm \sqrt{\lambda^2-1}\}$ and thus undergoes an exceptional-point-type transition at $\vert \lambda_\text{EP}^\pm \vert =1$. The spontaneous-symmetry-breaking phase transitions marked by these  EPs in the tangent flow are rooted in the local $\cPT$ symmetry of the linearization with time reversal $\cT$ ($t\rightarrow-t$) and parity reflection $\cP$ exchanging the species $y_1 \leftrightarrow y_3$:
\begin{equation}
\label{eq35}
    \cP = \begin{pmatrix}
        0 & 0 & 1 \\
        0 & 1 & 0 \\
        1 & 0 & 0
    \end{pmatrix} 
    \, , \qquad 
    \cP^2=\mathbbm{1}
    \, .
\end{equation}
This symmetry is furthermore a global symmetry of the nonlinear system, so that for any solution $\mathbf{y}$,
\newpage
\begin{equation}
\label{eq36}
    \mathbf{y}' = \cPT\mathbf{y} = \big( y_3(-t), y_2(-t), y_1(-t) \big)^T 
\end{equation}
is also a solution of the system.

As in Sec.~\ref{s5}, we can construct a Lyapunov function by considering the topologically equivalent four-dimensional replicator equation of (\ref{eq34}), for which the payoff matrix $A_\lambda$ is obtained from the intrinsic growth rates $\mathbf{r}$ and the interaction matrix $B$ as described in (\ref{eq20}).
We again use the freedom to add constant column contributions to this payoff matrix to eliminate its diagonal entries, without changing the dynamics.
The resulting matrix, 
\begin{equation}
    A_\lambda'=
    \frac{1}{\mathcal{N}}
    \begin{pmatrix}
    0 & - a & 0 & a \\
    1 & 0 & -a-1-\delta & a+\delta \\
    0 & 4-2a+\delta & 0 & 2a-4-\delta \\
    -1 & - a -\delta & 1+a+\delta & 0 
   \end{pmatrix}\\ 
\, ,
\end{equation}
with $a= 1+ \mathcal{N} \lambda/\sqrt{3}$,
is then antisymmetrizable by the diagonal matrix 
\begin{equation}
\label{eq37}
    D = d\cdot  \mathrm{diag}\Big(
    [1+\lambda\tfrac{\mathcal{N}}{\sqrt{3}}], 1, \mfrac{[2+\delta-2\lambda\tfrac{\mathcal{N}}{\sqrt{3}}]}{[2+\delta +\lambda \tfrac{\mathcal{N}}{\sqrt{3}}]} ,1 
    \Big)
\end{equation}
as long as all entries are positive, compare (\ref{eq27}).  For $\delta \in \{0,\pm1\}$, this is ensured between $\lambda = -\frac{\sqrt{3}}{\mathcal{N}}$ and 
$\lambda = (1+\frac{\delta}{2}) \frac{\sqrt{3}}{\mathcal{N}}$.
This will restrict the regime of deformation strengths $\lambda$ in which this approach results in a global Lyapunov function, as before.
The resulting constant of motion based on (\ref{eq37}) with $d=[2+\delta +\lambda \tfrac{\mathcal{N}}{\sqrt{3}}]\,/\,[2+\delta-2\lambda\tfrac{\mathcal{N}}{\sqrt{3}}]$ 
has the form
\begin{widetext}
\begin{equation}
    C(\mathbf{x}, \lambda) =
    \frac{
        \big[ 
        x_0  
        \big]^{ 
            \big(
            \frac{1}{1+\lambda\frac{\mathcal{N}}{\sqrt{3}}}
            \big)
                \big/
            \mathcal{M}
        }  
        \big[ 
        x_2 
        \big]^{ 
            \big(
            \frac{
            [2+\delta+\lambda\frac{\mathcal{N}}{\sqrt{3}}]}{
            [2+\delta -2\lambda \frac{\mathcal{N}}{\sqrt{3}}]}
            \big)
                \big/
            \mathcal{M}
        }
    }{
        \Big[
        x_1+ x_3 
        +x_2 \big
            (\frac{[2+\delta+\lambda\frac{\mathcal{N}}{\sqrt{3}}]}{[2+\delta -2\lambda \frac{\mathcal{N}}{\sqrt{3}}]}
            \big) 
        +x_0 \big(\frac{1}{1+\lambda\frac{\mathcal{N}}{\sqrt{3}}}\big)
        \Big]
        \Big(
            \frac{[2+\delta -2\lambda \frac{\mathcal{N}}{\sqrt{3}}]}{
            [2+\delta+\lambda\frac{\mathcal{N}}{\sqrt{3}}]}
        \Big) 
    }
    \,\,
    \Bigg[ 
    x_1 x_3 
    \Bigg(
        \mfrac{[2+\delta-2\lambda\tfrac{\mathcal{N}}{\sqrt{3}}]}{
        [2+\delta +\lambda \tfrac{\mathcal{N}}{\sqrt{3}}]}
    \Bigg)^2 \,
    \Bigg]^{ \, 
        1 \big/
        \mathcal{M}
    }
\! ,
\end{equation}
where $\mathcal{M} = 
    \frac{1}{1+\lambda\frac{\mathcal{N}}{\sqrt{3}}} 
    +3 \frac{
        [2+\delta-\lambda\frac{\mathcal{N}}{\sqrt{3}}]}{
        [2+\delta -2\lambda \frac{\mathcal{N}}{\sqrt{3}}]}$.
After dropping the $x_0$ component and mapping to the GLV, this then gives rise to the real-valued energy-like function
\begin{equation}
    H(\mathbf{y}, \lambda) = 
        \frac{
        \big[ 
        y_2 
        \big]^{ 
            \big(
            \frac{
            [2+\delta+\lambda\frac{\mathcal{N}}{\sqrt{3}}]}{
            [2+\delta -2\lambda \frac{\mathcal{N}}{\sqrt{3}}]}
            \big)
                \big/
            \mathcal{M}
        }
    }{
        y_1+ y_3 
        +y_2 \big
            (\frac{[2+\delta+\lambda\frac{\mathcal{N}}{\sqrt{3}}]}{[2+\delta -2\lambda \frac{\mathcal{N}}{\sqrt{3}}]}
            \big) 
        +\big(\frac{1}{1+\lambda\frac{\mathcal{N}}{\sqrt{3}}}\big)
    }
    \,\,
    \Bigg[ 
    y_1 y_3 \,
    \Bigg(
        \mfrac{[2+\delta-2\lambda\tfrac{\mathcal{N}}{\sqrt{3}}]}{
        [2+\delta +\lambda \tfrac{\mathcal{N}}{\sqrt{3}}]}
    \Bigg)^2 \,
    \Bigg]^{ \, 
        1 \big/
        \mathcal{M}
    }
\! .
\end{equation}
\end{widetext}
The function  $L(\mathbf{y},\lambda) = H( \mathbf{1} ,\lambda)- H(\mathbf{y},\lambda)$ then provides a Lyapunov function that ensures the marginal stability of the coexistence point and thus the concurrence of the linearized behavior with the nonlinear dynamics in its vicinity for $\vert\lambda\vert < \vert \lambda_\mathrm{EP}^\pm \vert = 1$. 
As in the two-dimensional model, the system is stable globally for small deformations $\lambda$, continuously transitions to a regime of regional stability as indicated by the breakdown of the antisymmetrizability of (\ref{eq37}), and becomes globally unstable for deformation strengths $\vert\lambda\vert > \vert \lambda_\mathrm{EP}^\pm \vert = 1$. 
In this, it replicates the correlation between EPs in the linearized local dynamics of a globally symmetric model and the global destabilization of the nonlinear dynamics in the context of a higher-dimensional system.
We illustrate this behavior for the food chain ($\delta=0$), the cyclic food chain ($\delta =1$), and the food chain with omnivory ($\delta=-1$) through the dynamics compacted to the $3$-simplex.

\subsection*{Food chain ($\delta=0$)}

The arguably simplest tri-trophic model is the \emph{food chain}, in which a primary producer species is preyed upon by an intermediate (meso-)predator, which in turn is the prey of an apex predator species that does not have a natural predator itself, cf. Fig.~\ref{f6a}. 
Here we consider the food chain model given by (\ref{eq34}) with $\delta=0$,
so that the dynamics are governed by the system 
\begin{equation}
\label{eq35}
    \begin{alignedat}{2}
    & \partial_t y_1 = \mfrac{y_1}{\mathcal{N}} [ 1 + (1+\lambda\tfrac{\mathcal{N}}{\sqrt{3}}) y_1 - (2+\lambda\tfrac{\mathcal{N}}{\sqrt{3}}) y_2 ] ,\\
    & \partial_t y_2 = \mfrac{y_2}{\mathcal{N}} [(3-\lambda\tfrac{\mathcal{N}}{\sqrt{3}}) y_1 + (\lambda\tfrac{\mathcal{N}}{\sqrt{3}} -3) y_3] ,\\
    & \partial_t y_3 = \mfrac{y_3}{\mathcal{N}} [-1 + (2+\lambda\tfrac{\mathcal{N}}{\sqrt{3}}) y_2 - (1+\lambda\tfrac{\mathcal{N}}{\sqrt{3}}) y_3]  ,
    \end{alignedat}
\end{equation}
with $\mathcal{N}=\sqrt{11}$.
Note that without deformation $\lambda$, the consumption rate of $y_1$ by $y_2$ and of $y_2$ by $y_3$ does not correspond to an equal sustenance rate of $y_2$ by $y_1$ and of $y_3$ by $y_2$. The deformation $\lambda$ adjusts this difference in the model. 

The behavior of the system is visualized in the $3$-simplex in the left column of Fig.~\ref{f7}.
Without deformation ($\lambda=0$) and at small deformation strengths, the system is globally stable, with periodic orbits indicated in blue, as reflected in the existence of a global Lyapunov function between 
$ -\sqrt{3/11} < \lambda < \sqrt{3/11} \approx 0.52$.
Thereafter, global stability breaks down to regional stability around the coexistence equilibrium, before the system abruptly becomes globally unstable at $\vert \lambda \vert > \vert \lambda_\mathrm{EP}^\pm \vert =1$. 
Trajectories moving toward sinks on the boundary of the system, corresponding to the eventual extinction of at least one species, are shown in red.

\subsection*{Cyclic food chain ($\delta=1$)}

When, instead of a linear chain with a primary producer and an apex predator species, each species preys 
upon one and is the prey of the other species in the three-dimensional system, it models cyclic competition, cf. Fig.~\ref{f6b}. 
For the cyclic food chain model given by (\ref{eq34}) with $\delta=1$,
the dynamics are described by
\begin{equation}
\label{eq35}
    \begin{alignedat}{2}
    & \partial_t y_1 = \mfrac{y_1}{\mathcal{N}} [ 1 + (1+\lambda\tfrac{\mathcal{N}}{\sqrt{3}}) y_1 - (3+\lambda\tfrac{\mathcal{N}}{\sqrt{3}}) y_2 +  y_3] ,\\
    & \partial_t y_2 = \mfrac{y_2}{\mathcal{N}} [(4-\lambda\tfrac{\mathcal{N}}{\sqrt{3}}) y_1 + (\lambda\tfrac{\mathcal{N}}{\sqrt{3}} -4) y_3] ,\\
    & \partial_t y_3 = \mfrac{y_3}{\mathcal{N}} [-1 -y_1 +(3+\lambda\tfrac{\mathcal{N}}{\sqrt{3}}) y_2 - (1+\lambda\tfrac{\mathcal{N}}{\sqrt{3}}) y_3] ,
    \end{alignedat}
\end{equation}
with $\mathcal{N}=2\sqrt{6}$.
As before, the deformation $\lambda$ adjusts this difference between the 
consumption rate of $y_1$ by $y_2$ ($y_2$ by $y_3$) and  
the sustenance rate of $y_2$ by $y_1$ ($y_3$ by $y_2$). 

The behavior of the system is visualized in the $3$-simplex in the right column of Fig.~\ref{f7}.
Again, the system is globally stable without deformation ($\lambda=0$) and at small deformation strengths, 
which is also mirrored in the existence of a global Lyapunov function between 
$ -0.35 \approx -\sqrt{2}/4 < \lambda < 3\sqrt{2}/8 \approx 0.53$.
Periodic orbits are indicated in blue.
Global stability then reduces continuously to regional stability around the coexistence equilibrium, before the system abruptly becomes globally unstable at $\vert \lambda \vert > \vert \lambda_\mathrm{EP}^\pm \vert =1$. 
In Fig.~\ref{f7}, trajectories in the unstable regions, resulting in the eventual extinction of at least one species, are shown in red. 


\setcounter{figure}{7}
\begin{figure}[H]
\centering
\subfloat[\centering $\lambda = 1.5, \delta=0$]{
\includegraphics[width=0.43\columnwidth]
{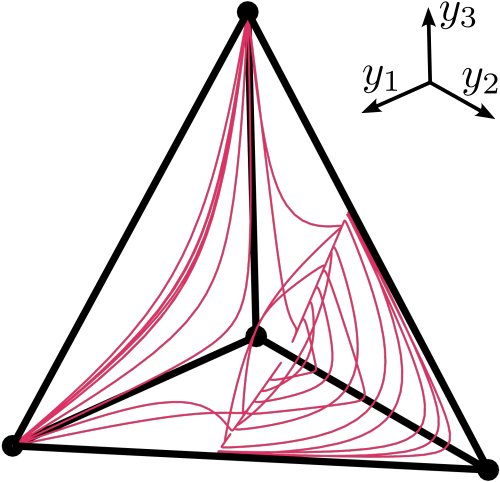}
\label{f7a}
}
\subfloat[\centering $\lambda = 1.5, \delta=1$]{
\includegraphics[width=0.43\columnwidth]
{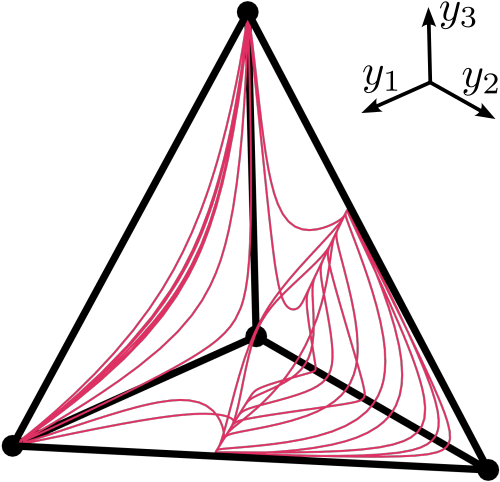}
\label{f7b}
}
\\[-1pt]
\subfloat[\centering $\lambda = 0.75, \delta=0$]{
\includegraphics[width=0.43\columnwidth]
{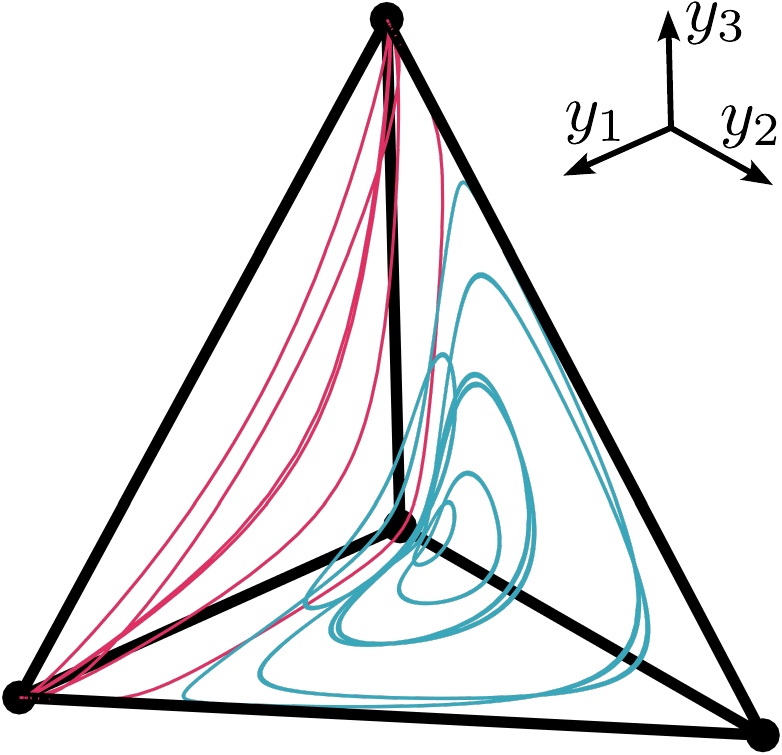}
\label{f7c}
}
\subfloat[\centering $\lambda = 0.75, \delta=1$]{
\includegraphics[width=0.43\columnwidth]
{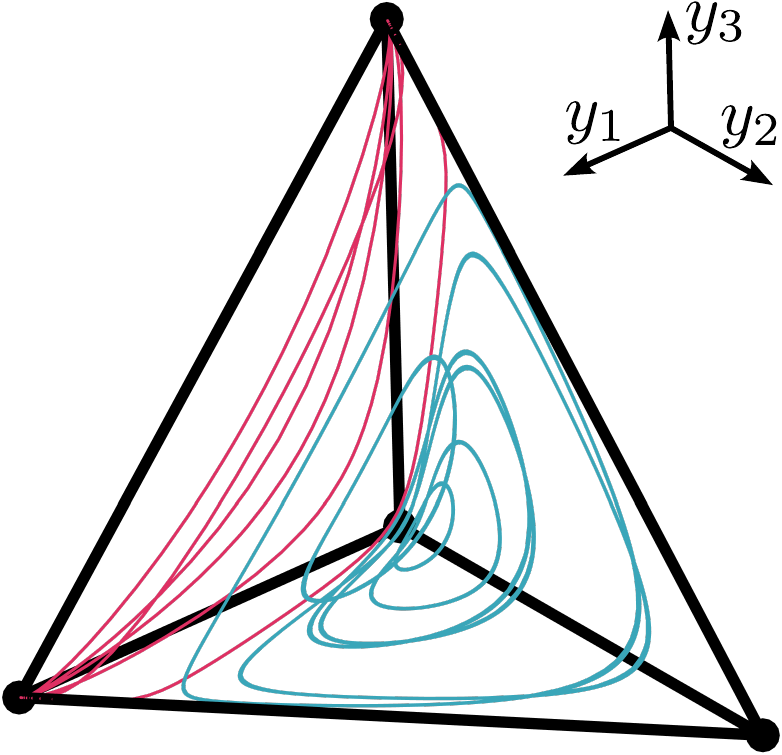}
\label{f7d}
}
\\[-1pt]
\subfloat[\centering $\lambda = 0, \delta=0$]{
\includegraphics[width=0.43\columnwidth]
{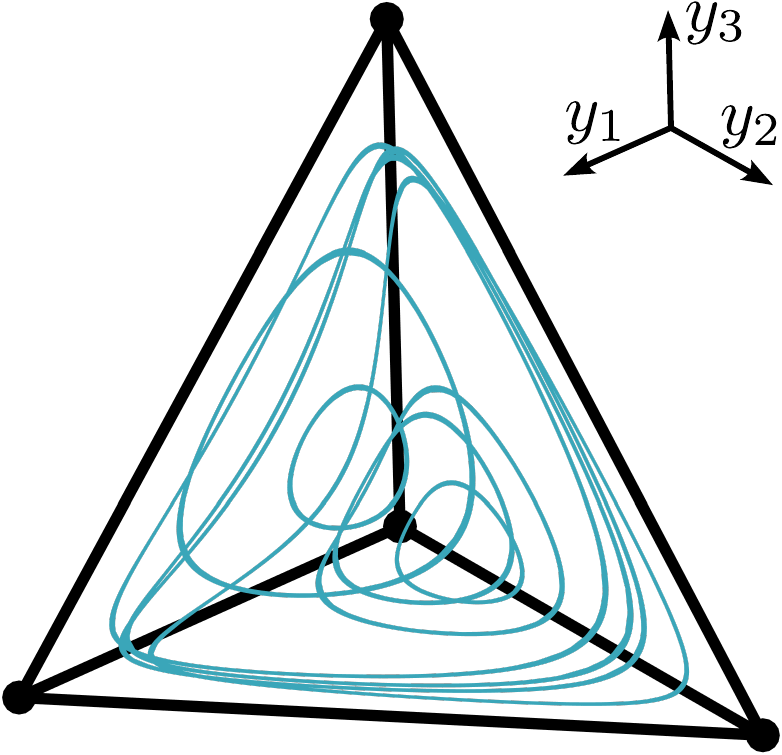}
\label{f7e}
}
\subfloat[\centering $\lambda = 0, \delta=1$]{
\includegraphics[width=0.43\columnwidth]
{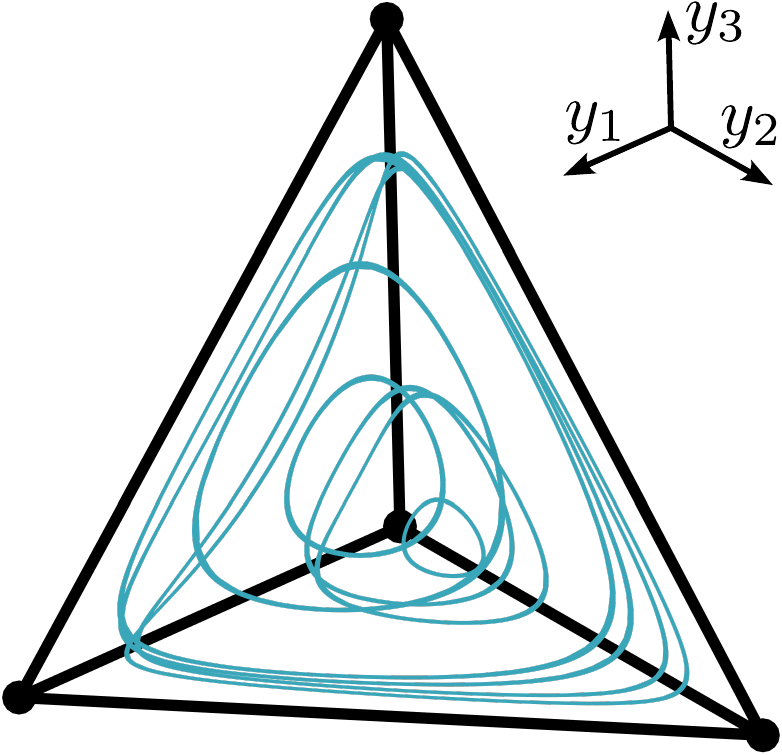}
\label{f7f}
}
\\[-1pt]
\subfloat[\centering $\lambda = -0.75, \delta=0$]{
\includegraphics[width=0.43\columnwidth]
{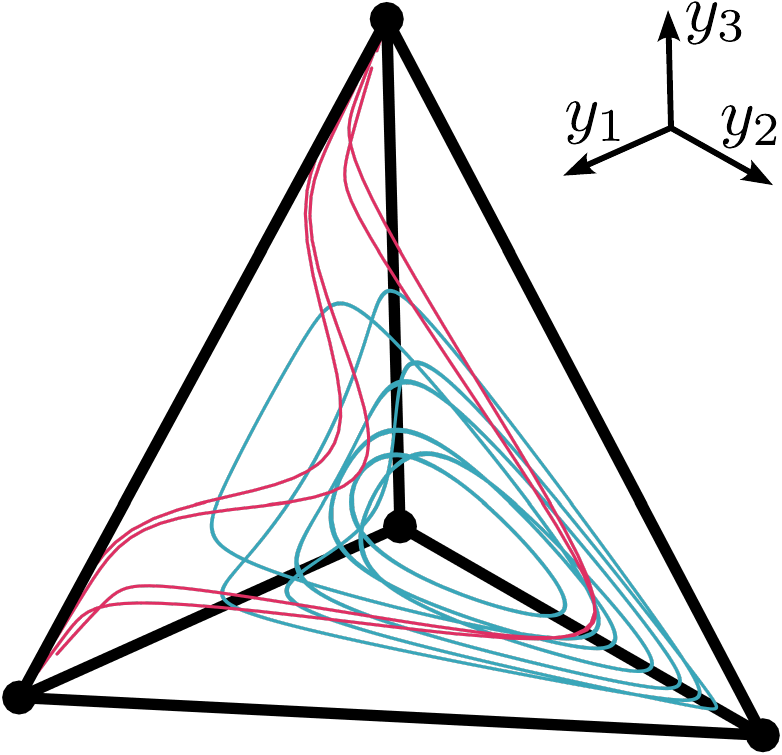}
\label{f7g}
}
\subfloat[\centering $\lambda = -0.75, \delta=1$]{
\includegraphics[width=0.43\columnwidth]
{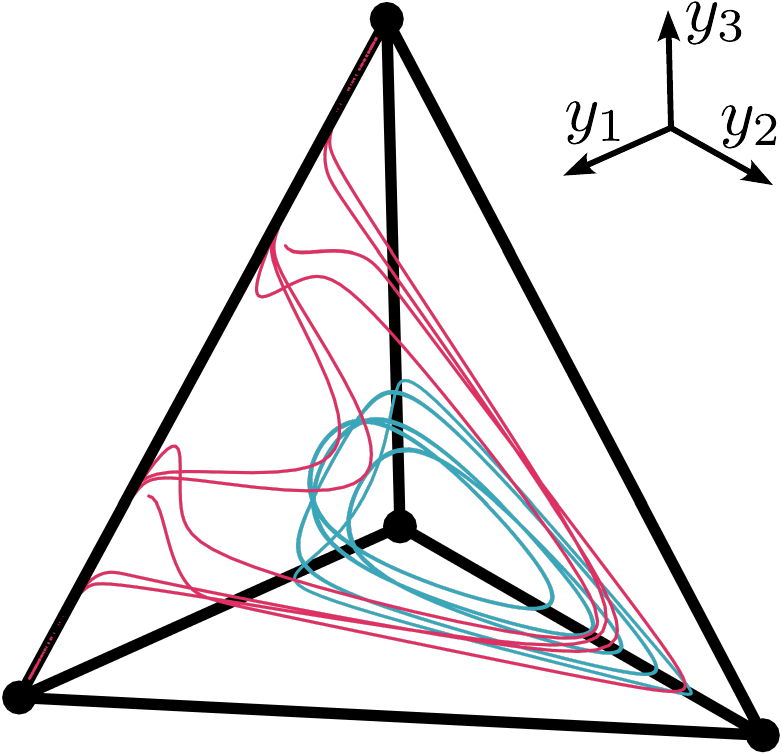}
\label{f7h}
}
\\[-1pt]
\subfloat[\centering $\lambda = -1.5, \delta=0$]{
\includegraphics[width=0.43\columnwidth]
{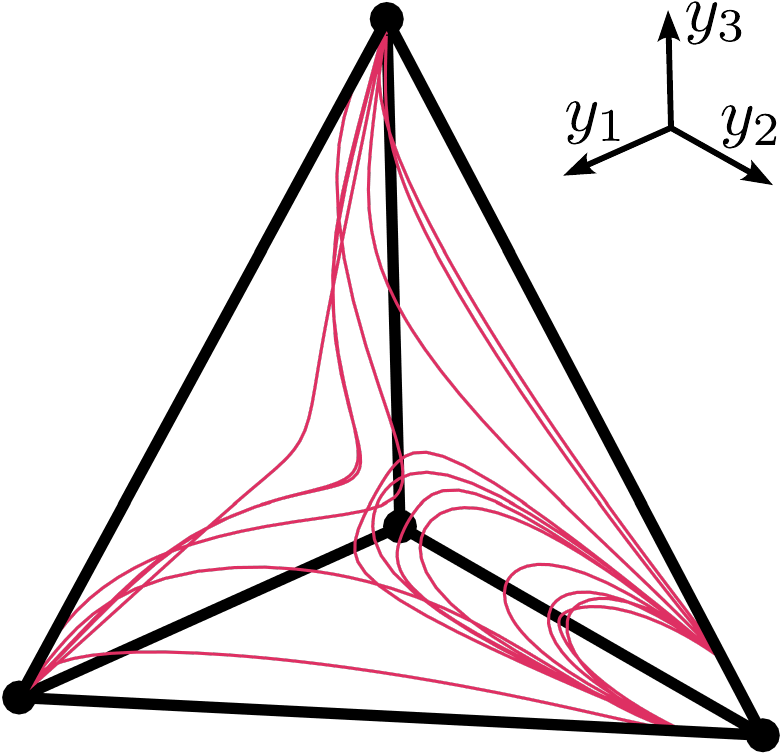}
\label{f7i}
}
\subfloat[\centering $\lambda = -1.5, \delta=1$]{
\includegraphics[width=0.43\columnwidth]
{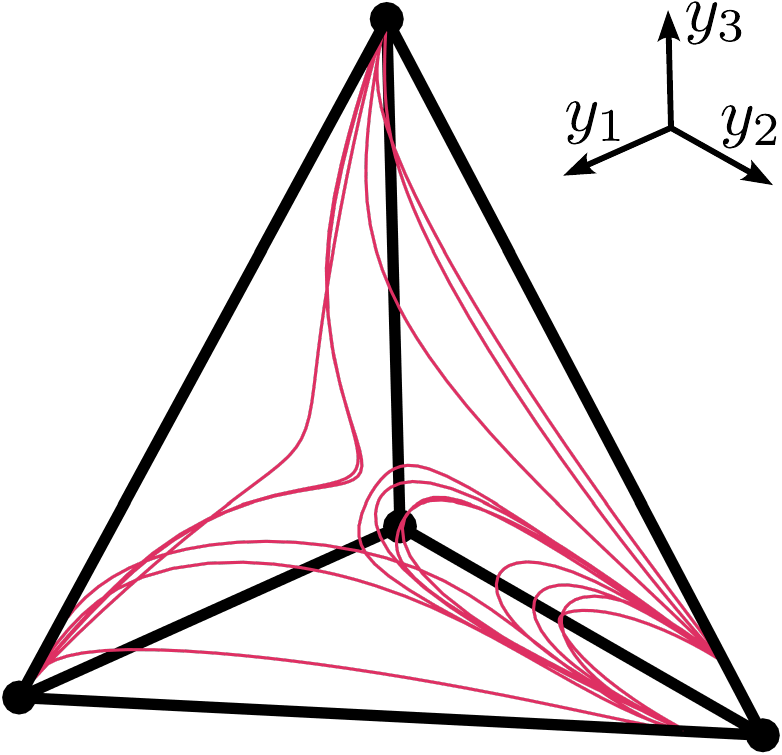}
\label{f7j}
}
\caption{
\label{f7}
Dynamics of the food-chain ($\delta=0$, left column) and cyclic-food-chain ($\delta=1$, right column) models on the $3$-simplex at exemplary values of the deformation value $\lambda$. Trajectories in the stable regime are shown in blue, trajectories evolving toward a boundary sink in red.
}
\end{figure}
\noindent

\subsection*{Food chain with omnivory ($\delta=-1$)}

Beyond the simple linear food chain, one may also consider generalist (omnivorous) predators, which prey upon multiple species. In the tri-trophic model, this means that both the primary producer species and the mesopredator species are preyed upon by the apex predators species, cf. Fig.~\ref{f6c}.
We consider the food chain model with omnivory given by (\ref{eq34}) with $\delta=-1$,
for which the 
dynamics are governed by
\begin{equation}
\label{eq35}
    \begin{alignedat}{2}
    & \partial_t y_1 = \mfrac{y_1}{\mathcal{N}} [ 1 + (1+\lambda\tfrac{\mathcal{N}}{\sqrt{3}}) y_1 - (1+\lambda\tfrac{\mathcal{N}}{\sqrt{3}}) y_2 -  y_3] ,\\
    & \partial_t y_2 = \mfrac{y_2}{\mathcal{N}} [(2-\lambda\tfrac{\mathcal{N}}{\sqrt{3}}) y_1 + (\lambda\tfrac{\mathcal{N}}{\sqrt{3}} -2) y_3] ,\\
    & \partial_t y_3 = \mfrac{y_3}{\mathcal{N}} [-1 + y_1 +(1+\lambda\tfrac{\mathcal{N}}{\sqrt{3}}) y_2 - (1+\lambda\tfrac{\mathcal{N}}{\sqrt{3}}) y_3] , 
    \end{alignedat}
\end{equation}
with $\mathcal{N}=2$.
The deformation $\lambda$ adjusts this difference between the 
consumption rate of $y_1$ by $y_2$ ($y_2$ by $y_3$) and  
the sustenance rate of $y_2$ by $y_1$ ($y_3$ by $y_2$). 

Figure \ref{f8} shows the behavior of the system in the $3$-simplex, with periodic orbits indicated in blue and trajectories moving toward a sink on the boundary, i.e., an eventual extinction of at least one species, in red.
Once again, the system is globally stable without deformation ($\lambda=0$) and at small deformation strengths, 
reflected in the existence of a global Lyapunov function between $ -0.87 \approx -\sqrt{3}/2 < \lambda < \sqrt{3}/4 \approx 0.43$.
This transitions continuously toward regional stability around the coexistence equilibrium and then abruptly becomes globally unstable at $\vert \lambda \vert > \vert \lambda_\mathrm{EP}^\pm \vert =1$.

\section{Conclusion}
\label{s7}
\setcounter{figure}{8}
\begin{figure*}[t]
\centering
\subfloat[\centering $\lambda = -1.5, \delta=-1$]{
\includegraphics[width=0.19\textwidth]
{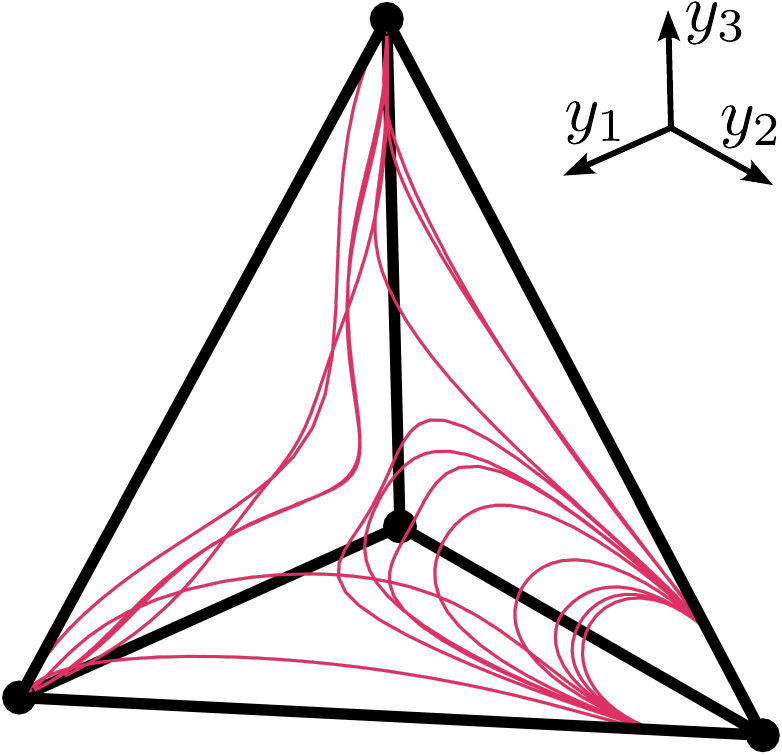}
\label{f8a}
}
\subfloat[\centering $\lambda = -0.9, \delta=-1$]{
\includegraphics[width=0.19\textwidth]
{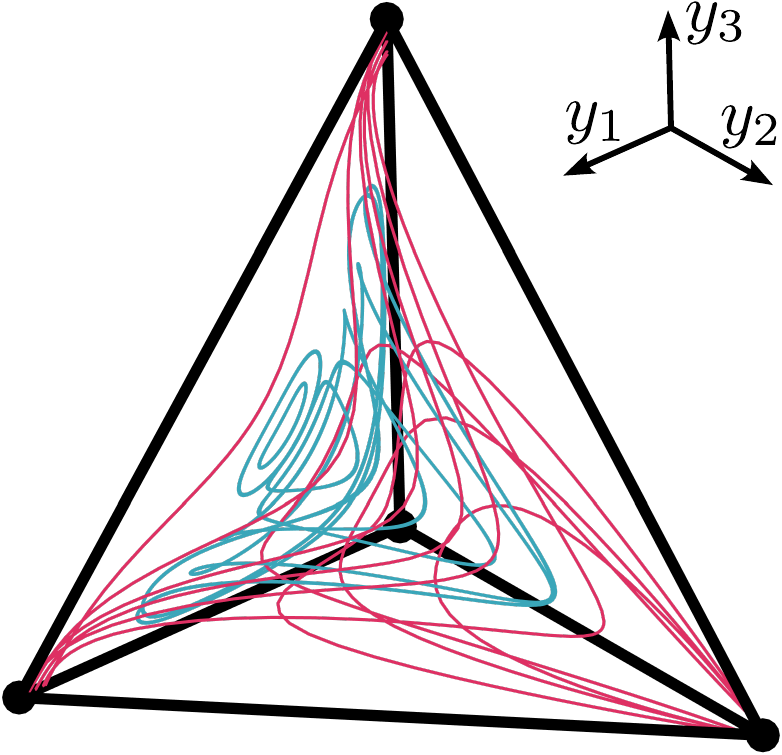}
\label{f8b}
}
\subfloat[\centering $\lambda = 0, \delta=-1$]{
\includegraphics[width=0.19\textwidth]
{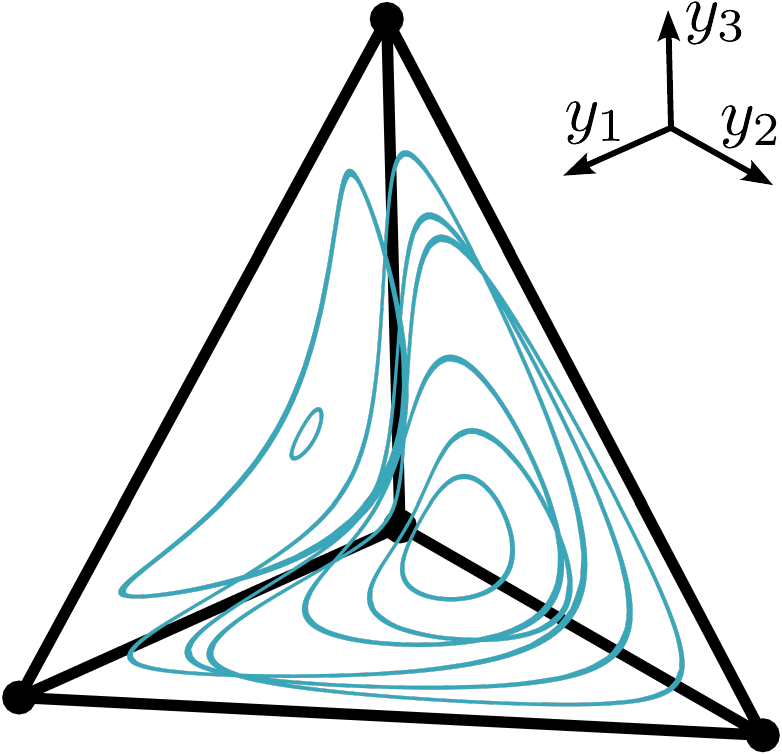}
\label{f8d}
}
\subfloat[\centering $\lambda = 0.75, \delta=-1$]{
\includegraphics[width=0.19\textwidth]
{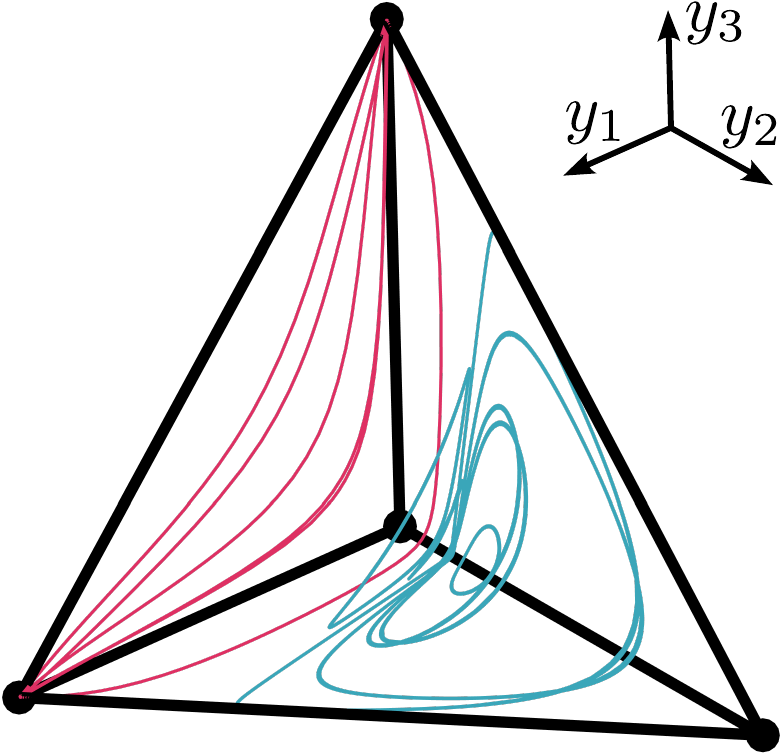}
\label{f8c}
}
\subfloat[\centering $\lambda = 1.5, \delta=-1$]{
\includegraphics[width=0.19\textwidth]
{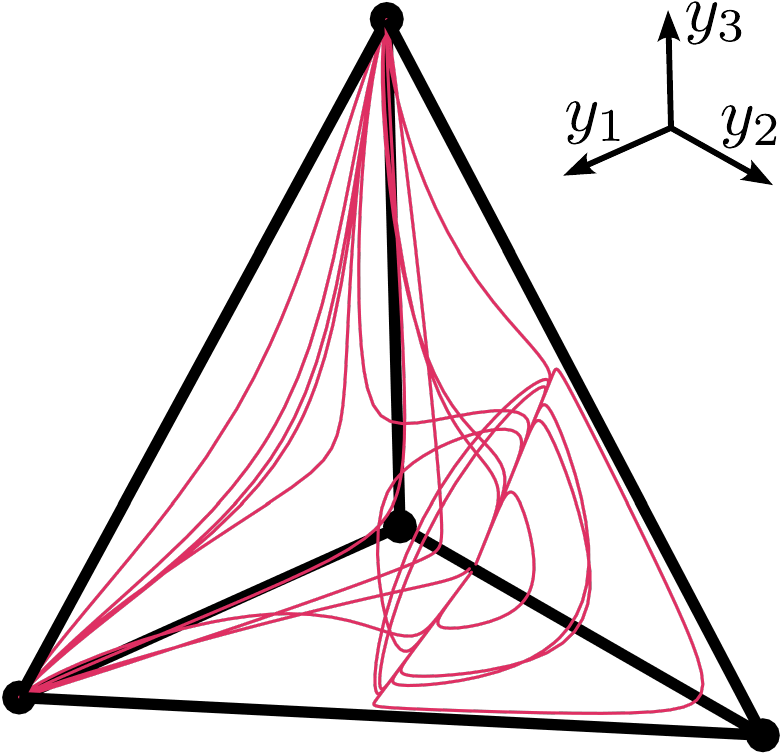}
\label{f8d}
}
\caption{
\label{f8}
Dynamics of the food chain with omnivory ($\delta=-1$) on the $3$-simplex at exemplary values of the deformation strength $\lambda$. Trajectories in the stable regime are shown in blue, trajectories evolving toward a boundary sink in red.
}
\end{figure*}

In this study, we have examined the well-established replicator dynamics of the rock-paper-scissors game and the generalized Lotka-Volterra system, being foundational models of evolutionary game theory and population ecology, through the lens of non-Hermitian physics. 
When linearized around their coexistence equilibrium, these systems are linked to common non-Hermitian models, undergoing spontaneous-symmetry-breaking phase transitions marked by EPs.
We have thereby outlined the restriction of this approach to linking nonlinearity and non-Hermiticity: 
In general, the linear non-Hermitian theory arises as a local description at the equilibrium only and is a priori not connected to the dynamics of the nonlinear system beyond it. 
In particular, the underlying symmetry, in which the non-Hermitian phase transition behavior and the occurrence of an EP is rooted, can emerge as a local feature of the equilibrium and does not have to be connected to global symmetries of the full nonlinear theory.
Moreover, mirroring a non-Hermitian $\cPT$-symmetric Sch\"odinger equation in the linearized dynamics of a nonlinear system commonly gives rise to phases of purely imaginary eigenvalues, such that the stationary state is a non-hyperbolic equilibrium. The validity of the linearization as a representative of the underlying nonlinear dynamics is thus not ensured in these cases, potentially disconnecting the linear and nonlinear model completely.

Nonetheless, when this connection is intact and the local symmetry of the linearized model is inherited from a global symmetry of the nonlinear system, the well-known phase-transition behavior of $\cPT$ theories may be indicative of global changes in the nonlinear dynamics. We have examined this connection in the context of the two- and three-dimensional Lotka-Volterra system, demonstrating an agreement of the EP with an abrupt global destabilization of the dynamics, in the sense that the system evolves toward a boundary state in which at least one species becomes extinct. 
With such systems being widely-used foundational models that capture central dynamical qualities in many areas of science, exceptional point structures may offer a new and valuable local characteristic for the understanding of nonlinear dynamics.

\section*{Acknowledgments}
We acknowledge funding from the Max Planck Society's Lise Meitner Excellence Program 2.0.

\vfill

\bibliography{references.bib}

\begin{thebibliography}{76}%
\makeatletter
\providecommand \@ifxundefined [1]{%
 \@ifx{#1\undefined}
}%
\providecommand \@ifnum [1]{%
 \ifnum #1\expandafter \@firstoftwo
 \else \expandafter \@secondoftwo
 \fi
}%
\providecommand \@ifx [1]{%
 \ifx #1\expandafter \@firstoftwo
 \else \expandafter \@secondoftwo
 \fi
}%
\providecommand \natexlab [1]{#1}%
\providecommand \enquote  [1]{``#1''}%
\providecommand \bibnamefont  [1]{#1}%
\providecommand \bibfnamefont [1]{#1}%
\providecommand \citenamefont [1]{#1}%
\providecommand \href@noop [0]{\@secondoftwo}%
\providecommand \href [0]{\begingroup \@sanitize@url \@href}%
\providecommand \@href[1]{\@@startlink{#1}\@@href}%
\providecommand \@@href[1]{\endgroup#1\@@endlink}%
\providecommand \@sanitize@url [0]{\catcode `\\12\catcode `\$12\catcode `\&12\catcode `\#12\catcode `\^12\catcode `\_12\catcode `\%12\relax}%
\providecommand \@@startlink[1]{}%
\providecommand \@@endlink[0]{}%
\providecommand \url  [0]{\begingroup\@sanitize@url \@url }%
\providecommand \@url [1]{\endgroup\@href {#1}{\urlprefix }}%
\providecommand \urlprefix  [0]{URL }%
\providecommand \Eprint [0]{\href }%
\providecommand \doibase [0]{https://doi.org/}%
\providecommand \selectlanguage [0]{\@gobble}%
\providecommand \bibinfo  [0]{\@secondoftwo}%
\providecommand \bibfield  [0]{\@secondoftwo}%
\providecommand \translation [1]{[#1]}%
\providecommand \BibitemOpen [0]{}%
\providecommand \bibitemStop [0]{}%
\providecommand \bibitemNoStop [0]{.\EOS\space}%
\providecommand \EOS [0]{\spacefactor3000\relax}%
\providecommand \BibitemShut  [1]{\csname bibitem#1\endcsname}%
\let\auto@bib@innerbib\@empty
\bibitem [{\citenamefont {Boyd}(2008)}]{B2008}%
  \BibitemOpen
  \bibfield  {author} {\bibinfo {author} {\bibfnamefont {R.~W.}\ \bibnamefont {Boyd}},\ }\href@noop {} {\emph {\bibinfo {title} {Nonlinear Optics}}}\ (\bibinfo  {publisher} {Academic Press},\ \bibinfo {year} {2008})\BibitemShut {NoStop}%
\bibitem [{\citenamefont {Besse}\ and\ \citenamefont {Garreau}(2015)}]{BG2015}%
  \BibitemOpen
  \bibfield  {author} {\bibinfo {author} {\bibfnamefont {C.}~\bibnamefont {Besse}}\ and\ \bibinfo {author} {\bibfnamefont {J.-C.}\ \bibnamefont {Garreau}},\ }\href {https://doi.org/10.1007/978-3-319-19015-0} {\emph {\bibinfo {title} {Nonlinear Optical and Atomic Systems}}}\ (\bibinfo  {publisher} {Springer Cham},\ \bibinfo {year} {2015})\BibitemShut {NoStop}%
\bibitem [{\citenamefont {Kirchgassner}(1975)}]{K1975}%
  \BibitemOpen
  \bibfield  {author} {\bibinfo {author} {\bibfnamefont {K.}~\bibnamefont {Kirchgassner}},\ }\bibfield  {title} {\bibinfo {title} {Bifurcation in nonlinear hydrodynamic stability},\ }\href {https://doi.org/10.1137/1017072} {\bibfield  {journal} {\bibinfo  {journal} {SIAM Review}\ }\textbf {\bibinfo {volume} {17}},\ \bibinfo {pages} {652} (\bibinfo {year} {1975})}\BibitemShut {NoStop}%
\bibitem [{\citenamefont {Cartwright}(2020)}]{C2020}%
  \BibitemOpen
  \bibfield  {author} {\bibinfo {author} {\bibfnamefont {J.}~\bibnamefont {Cartwright}},\ }\bibfield  {title} {\bibinfo {title} {Nonlinear dynamics determines the thermodynamic instability of condensed matter in vacuo},\ }\href {https://doi.org/10.1098/rsta.2019.0534} {\bibfield  {journal} {\bibinfo  {journal} {Philos. Trans. R. Soc. A}\ }\textbf {\bibinfo {volume} {378}},\ \bibinfo {pages} {20190534} (\bibinfo {year} {2020})}\BibitemShut {NoStop}%
\bibitem [{\citenamefont {Strogatz}(2000)}]{S2000}%
  \BibitemOpen
  \bibfield  {author} {\bibinfo {author} {\bibfnamefont {S.~H.}\ \bibnamefont {Strogatz}},\ }\href {https://doi.org/10.1201/9780429492563} {\emph {\bibinfo {title} {Nonlinear Dynamics and Chaos: With Applications to Physics, Biology, Chemistry and Engineering}}}\ (\bibinfo  {publisher} {Westview Press},\ \bibinfo {year} {2000})\BibitemShut {NoStop}%
\bibitem [{\citenamefont {Clark}\ and\ \citenamefont {Luis}(2020)}]{CL2020}%
  \BibitemOpen
  \bibfield  {author} {\bibinfo {author} {\bibfnamefont {T.}~\bibnamefont {Clark}}\ and\ \bibinfo {author} {\bibfnamefont {A.}~\bibnamefont {Luis}},\ }\bibfield  {title} {\bibinfo {title} {Nonlinear population dynamics are ubiquitous in animals},\ }\href {https://doi.org/10.1038/s41559-019-1052-6} {\bibfield  {journal} {\bibinfo  {journal} {Nat. Ecol. Evol.}\ }\textbf {\bibinfo {volume} {4}},\ \bibinfo {pages} {1} (\bibinfo {year} {2020})}\BibitemShut {NoStop}%
\bibitem [{\citenamefont {Sinervo}\ and\ \citenamefont {Lively}(1996)}]{SL1996}%
  \BibitemOpen
  \bibfield  {author} {\bibinfo {author} {\bibfnamefont {B.}~\bibnamefont {Sinervo}}\ and\ \bibinfo {author} {\bibfnamefont {C.}~\bibnamefont {Lively}},\ }\bibfield  {title} {\bibinfo {title} {The rock-paper-scissors game and the evolution of alternative male strategies},\ }\href {https://doi.org/10.1038/380240a0} {\bibfield  {journal} {\bibinfo  {journal} {Nature}\ }\textbf {\bibinfo {volume} {380}},\ \bibinfo {pages} {240} (\bibinfo {year} {1996})}\BibitemShut {NoStop}%
\bibitem [{\citenamefont {May}\ and\ \citenamefont {Leonard}(1975)}]{ML1975}%
  \BibitemOpen
  \bibfield  {author} {\bibinfo {author} {\bibfnamefont {R.}~\bibnamefont {May}}\ and\ \bibinfo {author} {\bibfnamefont {W.}~\bibnamefont {Leonard}},\ }\bibfield  {title} {\bibinfo {title} {Nonlinear aspects of competition between three species},\ }\href {https://doi.org/10.1137/0129022} {\bibfield  {journal} {\bibinfo  {journal} {SIAM J. Appl. Math.}\ }\textbf {\bibinfo {volume} {29}},\ \bibinfo {pages} {243} (\bibinfo {year} {1975})}\BibitemShut {NoStop}%
\bibitem [{\citenamefont {Kerr}\ \emph {et~al.}(2002)\citenamefont {Kerr}, \citenamefont {Riley}, \citenamefont {Feldman},\ and\ \citenamefont {Bohannan}}]{KRF2002}%
  \BibitemOpen
  \bibfield  {author} {\bibinfo {author} {\bibfnamefont {B.}~\bibnamefont {Kerr}}, \bibinfo {author} {\bibfnamefont {M.}~\bibnamefont {Riley}}, \bibinfo {author} {\bibfnamefont {M.}~\bibnamefont {Feldman}},\ and\ \bibinfo {author} {\bibfnamefont {B.}~\bibnamefont {Bohannan}},\ }\bibfield  {title} {\bibinfo {title} {Local dispersal promotes biodiversity in a real game of rock-paper-scissors},\ }\href {https://doi.org/10.1038/nature00823} {\bibfield  {journal} {\bibinfo  {journal} {Nature}\ }\textbf {\bibinfo {volume} {418}},\ \bibinfo {pages} {171} (\bibinfo {year} {2002})}\BibitemShut {NoStop}%
\bibitem [{\citenamefont {Noonburg}(1989)}]{N1989}%
  \BibitemOpen
  \bibfield  {author} {\bibinfo {author} {\bibfnamefont {V.~W.}\ \bibnamefont {Noonburg}},\ }\bibfield  {title} {\bibinfo {title} {A neural network modeled by an adaptive lotka-volterra system},\ }\href {https://doi.org/10.1137/0149109} {\bibfield  {journal} {\bibinfo  {journal} {SIAM J. Appl. Math.}\ }\textbf {\bibinfo {volume} {49}},\ \bibinfo {pages} {1779} (\bibinfo {year} {1989})}\BibitemShut {NoStop}%
\bibitem [{\citenamefont {Whalen}\ \emph {et~al.}(2015)\citenamefont {Whalen}, \citenamefont {Brennan}, \citenamefont {Sauer},\ and\ \citenamefont {Schiff}}]{WBS2015}%
  \BibitemOpen
  \bibfield  {author} {\bibinfo {author} {\bibfnamefont {A.}~\bibnamefont {Whalen}}, \bibinfo {author} {\bibfnamefont {S.}~\bibnamefont {Brennan}}, \bibinfo {author} {\bibfnamefont {T.}~\bibnamefont {Sauer}},\ and\ \bibinfo {author} {\bibfnamefont {S.}~\bibnamefont {Schiff}},\ }\bibfield  {title} {\bibinfo {title} {Observability and controllability of nonlinear networks: The role of symmetry},\ }\href {https://doi.org/10.1103/PhysRevX.5.011005} {\bibfield  {journal} {\bibinfo  {journal} {Phys. Rev. X}\ }\textbf {\bibinfo {volume} {5}},\ \bibinfo {pages} {011005} (\bibinfo {year} {2015})}\BibitemShut {NoStop}%
\bibitem [{\citenamefont {Szabo}\ and\ \citenamefont {Fath}(2006)}]{SF2006}%
  \BibitemOpen
  \bibfield  {author} {\bibinfo {author} {\bibfnamefont {G.}~\bibnamefont {Szabo}}\ and\ \bibinfo {author} {\bibfnamefont {G.}~\bibnamefont {Fath}},\ }\bibfield  {title} {\bibinfo {title} {Evolutionary games on graphs},\ }\href {https://doi.org/10.1016/j.physrep.2007.04.004} {\bibfield  {journal} {\bibinfo  {journal} {Phys. Rep.}\ }\textbf {\bibinfo {volume} {446}},\ \bibinfo {pages} {97} (\bibinfo {year} {2006})}\BibitemShut {NoStop}%
\bibitem [{\citenamefont {Barnett}\ \emph {et~al.}(1997)\citenamefont {Barnett}, \citenamefont {Medio},\ and\ \citenamefont {Serletis}}]{BMS1997}%
  \BibitemOpen
  \bibfield  {author} {\bibinfo {author} {\bibfnamefont {W.}~\bibnamefont {Barnett}}, \bibinfo {author} {\bibfnamefont {A.}~\bibnamefont {Medio}},\ and\ \bibinfo {author} {\bibfnamefont {A.}~\bibnamefont {Serletis}},\ }\bibfield  {title} {\bibinfo {title} {Nonlinear and complex dynamics in economics},\ }\href {https://doi.org/10.1017/S1365100514000091} {\bibfield  {journal} {\bibinfo  {journal} {Macroecon. Dyn.}\ }\textbf {\bibinfo {volume} {19}},\ \bibinfo {pages} {1749} (\bibinfo {year} {1997})}\BibitemShut {NoStop}%
\bibitem [{\citenamefont {Hauert}\ \emph {et~al.}(2002)\citenamefont {Hauert}, \citenamefont {Monte}, \citenamefont {Hofbauer},\ and\ \citenamefont {Sigmund}}]{HMH2002}%
  \BibitemOpen
  \bibfield  {author} {\bibinfo {author} {\bibfnamefont {C.}~\bibnamefont {Hauert}}, \bibinfo {author} {\bibfnamefont {S.}~\bibnamefont {Monte}}, \bibinfo {author} {\bibfnamefont {J.}~\bibnamefont {Hofbauer}},\ and\ \bibinfo {author} {\bibfnamefont {K.}~\bibnamefont {Sigmund}},\ }\bibfield  {title} {\bibinfo {title} {Volunteering as red queen mechanism for cooperation in public goods games},\ }\href {https://doi.org/10.1126/science.1070582} {\bibfield  {journal} {\bibinfo  {journal} {Science}\ }\textbf {\bibinfo {volume} {296}},\ \bibinfo {pages} {1129} (\bibinfo {year} {2002})}\BibitemShut {NoStop}%
\bibitem [{\citenamefont {Tainaka}(1993)}]{T1993}%
  \BibitemOpen
  \bibfield  {author} {\bibinfo {author} {\bibfnamefont {K.}~\bibnamefont {Tainaka}},\ }\bibfield  {title} {\bibinfo {title} {Paradoxical effect in a three-candidate voter model},\ }\href {https://doi.org/10.1016/0375-9601(93)90923-N} {\bibfield  {journal} {\bibinfo  {journal} {Phys. Lett. A}\ }\textbf {\bibinfo {volume} {176}},\ \bibinfo {pages} {303} (\bibinfo {year} {1993})}\BibitemShut {NoStop}%
\bibitem [{\citenamefont {Goodwin}(1982)}]{G1982}%
  \BibitemOpen
  \bibfield  {author} {\bibinfo {author} {\bibfnamefont {R.~M.}\ \bibnamefont {Goodwin}},\ }\bibfield  {title} {\bibinfo {title} {A growth cycle},\ }in\ \href {https://doi.org/10.1007/978-1-349-05504-3_12} {\emph {\bibinfo {booktitle} {Essays in Economic Dynamics}}}\ (\bibinfo  {publisher} {Palgrave Macmillan UK},\ \bibinfo {year} {1982})\ pp.\ \bibinfo {pages} {165--170}\BibitemShut {NoStop}%
\bibitem [{\citenamefont {Thompson}\ and\ \citenamefont {Stewart}(2002)}]{TS2002}%
  \BibitemOpen
  \bibfield  {author} {\bibinfo {author} {\bibfnamefont {J.~M.~T.}\ \bibnamefont {Thompson}}\ and\ \bibinfo {author} {\bibfnamefont {H.~B.}\ \bibnamefont {Stewart}},\ }\href@noop {} {\emph {\bibinfo {title} {Nonlinear dynamics and chaos}}}\ (\bibinfo  {publisher} {John Wiley \& Sons},\ \bibinfo {year} {2002})\BibitemShut {NoStop}%
\bibitem [{\citenamefont {Bender}\ and\ \citenamefont {Boettcher}(1998)}]{BB1998}%
  \BibitemOpen
  \bibfield  {author} {\bibinfo {author} {\bibfnamefont {C.~M.}\ \bibnamefont {Bender}}\ and\ \bibinfo {author} {\bibfnamefont {S.}~\bibnamefont {Boettcher}},\ }\bibfield  {title} {\bibinfo {title} {Real spectra in non-hermitian hamiltonians having $\mathcal{P}\mathcal{T}$ symmetry},\ }\href {https://doi.org/10.1103/PhysRevLett.80.5243} {\bibfield  {journal} {\bibinfo  {journal} {Phys. Rev. Lett.}\ }\textbf {\bibinfo {volume} {80}},\ \bibinfo {pages} {5243} (\bibinfo {year} {1998})}\BibitemShut {NoStop}%
\bibitem [{\citenamefont {Rotter}(2009)}]{R2009}%
  \BibitemOpen
  \bibfield  {author} {\bibinfo {author} {\bibfnamefont {I.}~\bibnamefont {Rotter}},\ }\bibfield  {title} {\bibinfo {title} {A non-hermitian hamilton operator and the physics of open quantum systems},\ }\href {https://doi.org/10.1088/1751-8113/42/15/153001} {\bibfield  {journal} {\bibinfo  {journal} {J. Phys. A: Math. Theor.}\ }\textbf {\bibinfo {volume} {42}},\ \bibinfo {pages} {153001} (\bibinfo {year} {2009})}\BibitemShut {NoStop}%
\bibitem [{\citenamefont {Roccati}\ \emph {et~al.}(2022)\citenamefont {Roccati}, \citenamefont {Palma}, \citenamefont {Ciccarello},\ and\ \citenamefont {Bagarello}}]{RPC2022}%
  \BibitemOpen
  \bibfield  {author} {\bibinfo {author} {\bibfnamefont {F.}~\bibnamefont {Roccati}}, \bibinfo {author} {\bibfnamefont {G.~M.}\ \bibnamefont {Palma}}, \bibinfo {author} {\bibfnamefont {F.}~\bibnamefont {Ciccarello}},\ and\ \bibinfo {author} {\bibfnamefont {F.}~\bibnamefont {Bagarello}},\ }\bibfield  {title} {\bibinfo {title} {Non-hermitian physics and master equations},\ }\href {https://doi.org/10.1142/S1230161222500044} {\bibfield  {journal} {\bibinfo  {journal} {Open Syst. Inf. Dyn.}\ }\textbf {\bibinfo {volume} {29}},\ \bibinfo {pages} {2250004} (\bibinfo {year} {2022})}\BibitemShut {NoStop}%
\bibitem [{\citenamefont {Bergholtz}\ \emph {et~al.}(2021)\citenamefont {Bergholtz}, \citenamefont {Budich},\ and\ \citenamefont {Kunst}}]{BBK2021}%
  \BibitemOpen
  \bibfield  {author} {\bibinfo {author} {\bibfnamefont {E.}~\bibnamefont {Bergholtz}}, \bibinfo {author} {\bibfnamefont {J.}~\bibnamefont {Budich}},\ and\ \bibinfo {author} {\bibfnamefont {F.}~\bibnamefont {Kunst}},\ }\bibfield  {title} {\bibinfo {title} {Exceptional topology of non-hermitian systems},\ }\href {https://doi.org/10.1103/RevModPhys.93.015005} {\bibfield  {journal} {\bibinfo  {journal} {Rev. Mod. Phys.}\ }\textbf {\bibinfo {volume} {93}},\ \bibinfo {pages} {015005} (\bibinfo {year} {2021})}\BibitemShut {NoStop}%
\bibitem [{\citenamefont {Kunst}\ \emph {et~al.}(2018)\citenamefont {Kunst}, \citenamefont {Edvardsson}, \citenamefont {Budich},\ and\ \citenamefont {Bergholtz}}]{KEB2018}%
  \BibitemOpen
  \bibfield  {author} {\bibinfo {author} {\bibfnamefont {F.}~\bibnamefont {Kunst}}, \bibinfo {author} {\bibfnamefont {E.}~\bibnamefont {Edvardsson}}, \bibinfo {author} {\bibfnamefont {J.}~\bibnamefont {Budich}},\ and\ \bibinfo {author} {\bibfnamefont {E.}~\bibnamefont {Bergholtz}},\ }\bibfield  {title} {\bibinfo {title} {Biorthogonal bulk-boundary correspondence in non-hermitian systems},\ }\href {https://doi.org/10.1103/PhysRevLett.121.026808} {\bibfield  {journal} {\bibinfo  {journal} {Phys. Rev. Lett.}\ }\textbf {\bibinfo {volume} {121}},\ \bibinfo {pages} {026808} (\bibinfo {year} {2018})}\BibitemShut {NoStop}%
\bibitem [{\citenamefont {Ding}\ \emph {et~al.}(2022)\citenamefont {Ding}, \citenamefont {Fang},\ and\ \citenamefont {Ma}}]{DFM2022}%
  \BibitemOpen
  \bibfield  {author} {\bibinfo {author} {\bibfnamefont {K.}~\bibnamefont {Ding}}, \bibinfo {author} {\bibfnamefont {C.}~\bibnamefont {Fang}},\ and\ \bibinfo {author} {\bibfnamefont {G.}~\bibnamefont {Ma}},\ }\bibfield  {title} {\bibinfo {title} {Non-hermitian topology and exceptional-point geometries},\ }\href {https://doi.org/10.1038/s42254-022-00516-5} {\bibfield  {journal} {\bibinfo  {journal} {Nat. Rev. Phys.}\ }\textbf {\bibinfo {volume} {4}},\ \bibinfo {pages} {745} (\bibinfo {year} {2022})}\BibitemShut {NoStop}%
\bibitem [{\citenamefont {Okuma}\ and\ \citenamefont {Sato}(2022)}]{OS2022}%
  \BibitemOpen
  \bibfield  {author} {\bibinfo {author} {\bibfnamefont {N.}~\bibnamefont {Okuma}}\ and\ \bibinfo {author} {\bibfnamefont {M.}~\bibnamefont {Sato}},\ }\bibfield  {title} {\bibinfo {title} {Non-hermitian topological phenomena: A review},\ }\href {https://doi.org/10.1146/annurev-conmatphys-040521-033133} {\bibfield  {journal} {\bibinfo  {journal} {Annu. Rev. Condens. Matter Phys.}\ }\textbf {\bibinfo {volume} {14}} (\bibinfo {year} {2022})}\BibitemShut {NoStop}%
\bibitem [{\citenamefont {Christodoulides}\ and\ \citenamefont {Yang}(2018)}]{CY2018}%
  \BibitemOpen
  \bibfield  {author} {\bibinfo {author} {\bibfnamefont {D.}~\bibnamefont {Christodoulides}}\ and\ \bibinfo {author} {\bibfnamefont {J.}~\bibnamefont {Yang}},\ }\href {https://doi.org/10.1007/978-981-13-1247-2} {\emph {\bibinfo {title} {Parity-time Symmetry and Its Applications}}}\ (\bibinfo  {publisher} {Springer Singapore},\ \bibinfo {year} {2018})\BibitemShut {NoStop}%
\bibitem [{\citenamefont {Charumathi}\ and\ \citenamefont {Senthilnathan}(2024)}]{CS2024}%
  \BibitemOpen
  \bibfield  {author} {\bibinfo {author} {\bibfnamefont {P.~R.}\ \bibnamefont {Charumathi}}\ and\ \bibinfo {author} {\bibfnamefont {K.}~\bibnamefont {Senthilnathan}},\ }\bibfield  {title} {\bibinfo {title} {Unravelling pt symmetry: Applications in metamaterials},\ }\href {https://doi.org/10.1007/s11468-024-02414-1} {\bibfield  {journal} {\bibinfo  {journal} {Plasmonics}\ ,\ \bibinfo {pages} {1}} (\bibinfo {year} {2024})}\BibitemShut {NoStop}%
\bibitem [{\citenamefont {Ge}\ and\ \citenamefont {El-Ganainy}(2016)}]{GE2016}%
  \BibitemOpen
  \bibfield  {author} {\bibinfo {author} {\bibfnamefont {L.}~\bibnamefont {Ge}}\ and\ \bibinfo {author} {\bibfnamefont {R.}~\bibnamefont {El-Ganainy}},\ }\bibfield  {title} {\bibinfo {title} {Nonlinear modal interactions in parity-time (pt) symmetric lasers},\ }\href {https://doi.org/10.1038/srep24889} {\bibfield  {journal} {\bibinfo  {journal} {Sci. Rep.}\ }\textbf {\bibinfo {volume} {6}},\ \bibinfo {pages} {24889} (\bibinfo {year} {2016})}\BibitemShut {NoStop}%
\bibitem [{\citenamefont {Ramezani}\ \emph {et~al.}(2010)\citenamefont {Ramezani}, \citenamefont {Kottos}, \citenamefont {El-Ganainy},\ and\ \citenamefont {Christodoulides}}]{RKE2010}%
  \BibitemOpen
  \bibfield  {author} {\bibinfo {author} {\bibfnamefont {H.}~\bibnamefont {Ramezani}}, \bibinfo {author} {\bibfnamefont {T.}~\bibnamefont {Kottos}}, \bibinfo {author} {\bibfnamefont {R.}~\bibnamefont {El-Ganainy}},\ and\ \bibinfo {author} {\bibfnamefont {D.}~\bibnamefont {Christodoulides}},\ }\bibfield  {title} {\bibinfo {title} {Unidirectional nonlinear pt-symmetric optical structures},\ }\href {https://doi.org/10.1103/PhysRevA.82.043803} {\bibfield  {journal} {\bibinfo  {journal} {Phys. Rev. A}\ }\textbf {\bibinfo {volume} {82}},\ \bibinfo {pages} {043803} (\bibinfo {year} {2010})}\BibitemShut {NoStop}%
\bibitem [{\citenamefont {Musslimani}\ \emph {et~al.}(2008)\citenamefont {Musslimani}, \citenamefont {Makris}, \citenamefont {El-Ganainy},\ and\ \citenamefont {Christodoulides}}]{MME2008}%
  \BibitemOpen
  \bibfield  {author} {\bibinfo {author} {\bibfnamefont {Z.}~\bibnamefont {Musslimani}}, \bibinfo {author} {\bibfnamefont {K.}~\bibnamefont {Makris}}, \bibinfo {author} {\bibfnamefont {R.}~\bibnamefont {El-Ganainy}},\ and\ \bibinfo {author} {\bibfnamefont {D.}~\bibnamefont {Christodoulides}},\ }\bibfield  {title} {\bibinfo {title} {Optical solitons in pt periodic potentials},\ }\href {https://doi.org/10.1103/PHYSREVLETT.100.030402} {\bibfield  {journal} {\bibinfo  {journal} {Phys. Rev. Lett.}\ }\textbf {\bibinfo {volume} {100}},\ \bibinfo {pages} {030402} (\bibinfo {year} {2008})}\BibitemShut {NoStop}%
\bibitem [{\citenamefont {Malzard}\ \emph {et~al.}(2018)\citenamefont {Malzard}, \citenamefont {Cancellieri},\ and\ \citenamefont {Schomerus}}]{MCS2018}%
  \BibitemOpen
  \bibfield  {author} {\bibinfo {author} {\bibfnamefont {S.}~\bibnamefont {Malzard}}, \bibinfo {author} {\bibfnamefont {E.}~\bibnamefont {Cancellieri}},\ and\ \bibinfo {author} {\bibfnamefont {H.}~\bibnamefont {Schomerus}},\ }\bibfield  {title} {\bibinfo {title} {Topological dynamics and excitations in lasers and condensates with saturable gain or loss},\ }\href {https://doi.org/10.1364/OE.26.022506} {\bibfield  {journal} {\bibinfo  {journal} {Opt. Express}\ }\textbf {\bibinfo {volume} {26}},\ \bibinfo {pages} {22506} (\bibinfo {year} {2018})}\BibitemShut {NoStop}%
\bibitem [{\citenamefont {Malzard}\ and\ \citenamefont {Schomerus}(2018)}]{MS2018}%
  \BibitemOpen
  \bibfield  {author} {\bibinfo {author} {\bibfnamefont {S.}~\bibnamefont {Malzard}}\ and\ \bibinfo {author} {\bibfnamefont {H.}~\bibnamefont {Schomerus}},\ }\bibfield  {title} {\bibinfo {title} {Nonlinear mode competition and symmetry-protected power oscillations in topological lasers},\ }\href {https://doi.org/10.1088/1367-2630/aac9e0} {\bibfield  {journal} {\bibinfo  {journal} {New J. Phys.}\ }\textbf {\bibinfo {volume} {20}},\ \bibinfo {pages} {063044} (\bibinfo {year} {2018})}\BibitemShut {NoStop}%
\bibitem [{\citenamefont {Lumer}\ \emph {et~al.}(2013)\citenamefont {Lumer}, \citenamefont {Plotnik}, \citenamefont {Rechtsman},\ and\ \citenamefont {Segev}}]{LPR2013}%
  \BibitemOpen
  \bibfield  {author} {\bibinfo {author} {\bibfnamefont {Y.}~\bibnamefont {Lumer}}, \bibinfo {author} {\bibfnamefont {Y.}~\bibnamefont {Plotnik}}, \bibinfo {author} {\bibfnamefont {M.~C.}\ \bibnamefont {Rechtsman}},\ and\ \bibinfo {author} {\bibfnamefont {M.}~\bibnamefont {Segev}},\ }\bibfield  {title} {\bibinfo {title} {Nonlinearly induced pt transition in photonic systems},\ }\href {https://doi.org/10.1103/PhysRevLett.111.263901} {\bibfield  {journal} {\bibinfo  {journal} {Phys. Rev. Lett.}\ }\textbf {\bibinfo {volume} {111}},\ \bibinfo {pages} {263901} (\bibinfo {year} {2013})}\BibitemShut {NoStop}%
\bibitem [{\citenamefont {Smirnova}\ \emph {et~al.}(2020)\citenamefont {Smirnova}, \citenamefont {Leykam}, \citenamefont {Chong},\ and\ \citenamefont {Kivshar}}]{SLC2020}%
  \BibitemOpen
  \bibfield  {author} {\bibinfo {author} {\bibfnamefont {D.}~\bibnamefont {Smirnova}}, \bibinfo {author} {\bibfnamefont {D.}~\bibnamefont {Leykam}}, \bibinfo {author} {\bibfnamefont {Y.}~\bibnamefont {Chong}},\ and\ \bibinfo {author} {\bibfnamefont {Y.}~\bibnamefont {Kivshar}},\ }\bibfield  {title} {\bibinfo {title} {Nonlinear topological photonics},\ }\href {https://doi.org/10.1063/1.5142397} {\bibfield  {journal} {\bibinfo  {journal} {Appl. Phys. Rev.}\ }\textbf {\bibinfo {volume} {7}},\ \bibinfo {pages} {021306} (\bibinfo {year} {2020})}\BibitemShut {NoStop}%
\bibitem [{\citenamefont {Ezawa}(2022)}]{E2022}%
  \BibitemOpen
  \bibfield  {author} {\bibinfo {author} {\bibfnamefont {M.}~\bibnamefont {Ezawa}},\ }\bibfield  {title} {\bibinfo {title} {Nonlinear non-hermitian higher-order topological laser},\ }\href {https://doi.org/10.1103/PhysRevResearch.4.013195} {\bibfield  {journal} {\bibinfo  {journal} {Phys. Rev. Res.}\ }\textbf {\bibinfo {volume} {4}},\ \bibinfo {pages} {013195} (\bibinfo {year} {2022})}\BibitemShut {NoStop}%
\bibitem [{\citenamefont {Yuce}(2021)}]{Y2021}%
  \BibitemOpen
  \bibfield  {author} {\bibinfo {author} {\bibfnamefont {C.}~\bibnamefont {Yuce}},\ }\bibfield  {title} {\bibinfo {title} {Nonlinear non-hermitian skin effect},\ }\href {https://doi.org/10.1016/j.physleta.2021.127484} {\bibfield  {journal} {\bibinfo  {journal} {Phys. Lett. A}\ }\textbf {\bibinfo {volume} {408}},\ \bibinfo {pages} {127484} (\bibinfo {year} {2021})}\BibitemShut {NoStop}%
\bibitem [{\citenamefont {Ma}\ and\ \citenamefont {Susanto}(2021)}]{MS2021}%
  \BibitemOpen
  \bibfield  {author} {\bibinfo {author} {\bibfnamefont {Y.-P.}\ \bibnamefont {Ma}}\ and\ \bibinfo {author} {\bibfnamefont {H.}~\bibnamefont {Susanto}},\ }\bibfield  {title} {\bibinfo {title} {Topological edge solitons and their stability in a nonlinear su-schrieffer-heeger model},\ }\href {https://doi.org/10.1103/PhysRevE.104.054206} {\bibfield  {journal} {\bibinfo  {journal} {Phys. Rev. E}\ }\textbf {\bibinfo {volume} {104}},\ \bibinfo {pages} {054206} (\bibinfo {year} {2021})}\BibitemShut {NoStop}%
\bibitem [{\citenamefont {Xia}\ \emph {et~al.}(2021)\citenamefont {Xia}, \citenamefont {Kaltsas}, \citenamefont {Song}, \citenamefont {Komis}, \citenamefont {Xu}, \citenamefont {Szameit}, \citenamefont {Buljan}, \citenamefont {Makris},\ and\ \citenamefont {Chen}}]{XKS2021}%
  \BibitemOpen
  \bibfield  {author} {\bibinfo {author} {\bibfnamefont {S.}~\bibnamefont {Xia}}, \bibinfo {author} {\bibfnamefont {D.}~\bibnamefont {Kaltsas}}, \bibinfo {author} {\bibfnamefont {D.}~\bibnamefont {Song}}, \bibinfo {author} {\bibfnamefont {I.}~\bibnamefont {Komis}}, \bibinfo {author} {\bibfnamefont {J.}~\bibnamefont {Xu}}, \bibinfo {author} {\bibfnamefont {A.}~\bibnamefont {Szameit}}, \bibinfo {author} {\bibfnamefont {H.}~\bibnamefont {Buljan}}, \bibinfo {author} {\bibfnamefont {K.}~\bibnamefont {Makris}},\ and\ \bibinfo {author} {\bibfnamefont {Z.}~\bibnamefont {Chen}},\ }\bibfield  {title} {\bibinfo {title} {Nonlinear tuning of pt symmetry and non-hermitian topological states},\ }\href {https://doi.org/10.1126/science.abf6873} {\bibfield  {journal} {\bibinfo  {journal} {Science}\ }\textbf {\bibinfo {volume} {372}},\ \bibinfo {pages} {72} (\bibinfo {year} {2021})}\BibitemShut {NoStop}%
\bibitem [{\citenamefont {Wimmer}\ \emph {et~al.}(2015)\citenamefont {Wimmer}, \citenamefont {Regensburger}, \citenamefont {Miri}, \citenamefont {Bersch}, \citenamefont {Christodoulides},\ and\ \citenamefont {Peschel}}]{WRM2015}%
  \BibitemOpen
  \bibfield  {author} {\bibinfo {author} {\bibfnamefont {M.}~\bibnamefont {Wimmer}}, \bibinfo {author} {\bibfnamefont {A.}~\bibnamefont {Regensburger}}, \bibinfo {author} {\bibfnamefont {M.-A.}\ \bibnamefont {Miri}}, \bibinfo {author} {\bibfnamefont {C.}~\bibnamefont {Bersch}}, \bibinfo {author} {\bibfnamefont {D.}~\bibnamefont {Christodoulides}},\ and\ \bibinfo {author} {\bibfnamefont {U.}~\bibnamefont {Peschel}},\ }\bibfield  {title} {\bibinfo {title} {Observation of optical solitons in pt-symmetric lattices},\ }\href {https://doi.org/10.1038/ncomms8782} {\bibfield  {journal} {\bibinfo  {journal} {Nat. Commun.}\ }\textbf {\bibinfo {volume} {6}},\ \bibinfo {pages} {7782} (\bibinfo {year} {2015})}\BibitemShut {NoStop}%
\bibitem [{\citenamefont {Haag}\ \emph {et~al.}(2014)\citenamefont {Haag}, \citenamefont {Dast}, \citenamefont {L{\"o}hle}, \citenamefont {Cartarius}, \citenamefont {Main},\ and\ \citenamefont {Wunner}}]{HDL2014}%
  \BibitemOpen
  \bibfield  {author} {\bibinfo {author} {\bibfnamefont {D.}~\bibnamefont {Haag}}, \bibinfo {author} {\bibfnamefont {D.}~\bibnamefont {Dast}}, \bibinfo {author} {\bibfnamefont {A.}~\bibnamefont {L{\"o}hle}}, \bibinfo {author} {\bibfnamefont {H.}~\bibnamefont {Cartarius}}, \bibinfo {author} {\bibfnamefont {J.}~\bibnamefont {Main}},\ and\ \bibinfo {author} {\bibfnamefont {G.}~\bibnamefont {Wunner}},\ }\bibfield  {title} {\bibinfo {title} {Nonlinear quantum dynamics in a pt-symmetric double well},\ }\href {https://doi.org/10.1103/PhysRevA.89.023601} {\bibfield  {journal} {\bibinfo  {journal} {Physical Review A}\ }\textbf {\bibinfo {volume} {89}},\ \bibinfo {pages} {023601} (\bibinfo {year} {2014})}\BibitemShut {NoStop}%
\bibitem [{\citenamefont {Graefe}\ \emph {et~al.}(2010)\citenamefont {Graefe}, \citenamefont {Korsch},\ and\ \citenamefont {Niederle}}]{GKN2010}%
  \BibitemOpen
  \bibfield  {author} {\bibinfo {author} {\bibfnamefont {E.-M.}\ \bibnamefont {Graefe}}, \bibinfo {author} {\bibfnamefont {H.~J.}\ \bibnamefont {Korsch}},\ and\ \bibinfo {author} {\bibfnamefont {A.~E.}\ \bibnamefont {Niederle}},\ }\bibfield  {title} {\bibinfo {title} {Quantum-classical correspondence for a non-hermitian bose-hubbard dimer},\ }\href {https://doi.org/10.1103/PhysRevA.82.013629} {\bibfield  {journal} {\bibinfo  {journal} {Physical Review A—Atomic, Molecular, and Optical Physics}\ }\textbf {\bibinfo {volume} {82}},\ \bibinfo {pages} {013629} (\bibinfo {year} {2010})}\BibitemShut {NoStop}%
\bibitem [{\citenamefont {Tang}\ \emph {et~al.}(2021)\citenamefont {Tang}, \citenamefont {Agudo-Canalejo},\ and\ \citenamefont {Golestanian}}]{TAG2021}%
  \BibitemOpen
  \bibfield  {author} {\bibinfo {author} {\bibfnamefont {E.}~\bibnamefont {Tang}}, \bibinfo {author} {\bibfnamefont {J.}~\bibnamefont {Agudo-Canalejo}},\ and\ \bibinfo {author} {\bibfnamefont {R.}~\bibnamefont {Golestanian}},\ }\bibfield  {title} {\bibinfo {title} {Topology protects chiral edge currents in stochastic systems},\ }\href {https://doi.org/10.1103/PhysRevX.11.031015} {\bibfield  {journal} {\bibinfo  {journal} {Phys. Rev. X}\ }\textbf {\bibinfo {volume} {11}},\ \bibinfo {pages} {031015} (\bibinfo {year} {2021})}\BibitemShut {NoStop}%
\bibitem [{\citenamefont {Nelson}\ and\ \citenamefont {Tang}(2024)}]{NT2024}%
  \BibitemOpen
  \bibfield  {author} {\bibinfo {author} {\bibfnamefont {A.}~\bibnamefont {Nelson}}\ and\ \bibinfo {author} {\bibfnamefont {E.}~\bibnamefont {Tang}},\ }\bibfield  {title} {\bibinfo {title} {Nonreciprocity is necessary for robust dimensional reduction and strong responses in stochastic topological systems},\ }\href {https://doi.org/10.1103/PhysRevB.110.155116} {\bibfield  {journal} {\bibinfo  {journal} {Phys. Rev. B}\ }\textbf {\bibinfo {volume} {110}},\ \bibinfo {pages} {155116} (\bibinfo {year} {2024})}\BibitemShut {NoStop}%
\bibitem [{\citenamefont {Zheng}\ and\ \citenamefont {Tang}(2024)}]{ZT2024}%
  \BibitemOpen
  \bibfield  {author} {\bibinfo {author} {\bibfnamefont {C.}~\bibnamefont {Zheng}}\ and\ \bibinfo {author} {\bibfnamefont {E.}~\bibnamefont {Tang}},\ }\bibfield  {title} {\bibinfo {title} {A topological mechanism for robust and efficient global oscillations in biological networks},\ }\href {https://doi.org/0.1038/s41467-024-50510-x} {\bibfield  {journal} {\bibinfo  {journal} {Nat. Commun.}\ }\textbf {\bibinfo {volume} {15}},\ \bibinfo {pages} {6453} (\bibinfo {year} {2024})}\BibitemShut {NoStop}%
\bibitem [{\citenamefont {Bender}\ \emph {et~al.}(2016)\citenamefont {Bender}, \citenamefont {Ghatak},\ and\ \citenamefont {Gianfreda}}]{BGG2016}%
  \BibitemOpen
  \bibfield  {author} {\bibinfo {author} {\bibfnamefont {C.~M.}\ \bibnamefont {Bender}}, \bibinfo {author} {\bibfnamefont {A.}~\bibnamefont {Ghatak}},\ and\ \bibinfo {author} {\bibfnamefont {M.}~\bibnamefont {Gianfreda}},\ }\bibfield  {title} {\bibinfo {title} {$\mathcal{P}\mathcal{T}$-symmetric model of immune response},\ }\href {https://doi.org/10.1088/1751-8121/50/3/035601} {\bibfield  {journal} {\bibinfo  {journal} {J. Phys. A Math.}\ }\textbf {\bibinfo {volume} {50}},\ \bibinfo {pages} {035601} (\bibinfo {year} {2016})}\BibitemShut {NoStop}%
\bibitem [{\citenamefont {Yoshida}\ \emph {et~al.}(2021)\citenamefont {Yoshida}, \citenamefont {Mizoguchi},\ and\ \citenamefont {Hatsugai}}]{YMH2021}%
  \BibitemOpen
  \bibfield  {author} {\bibinfo {author} {\bibfnamefont {T.}~\bibnamefont {Yoshida}}, \bibinfo {author} {\bibfnamefont {T.}~\bibnamefont {Mizoguchi}},\ and\ \bibinfo {author} {\bibfnamefont {Y.}~\bibnamefont {Hatsugai}},\ }\bibfield  {title} {\bibinfo {title} {Chiral edge modes in evolutionary game theory: A kagome network of rock-paper-scissors cycles},\ }\href {https://doi.org/10.1103/PhysRevE.104.025003} {\bibfield  {journal} {\bibinfo  {journal} {Phys. Rev. E}\ }\textbf {\bibinfo {volume} {104}},\ \bibinfo {pages} {025003} (\bibinfo {year} {2021})}\BibitemShut {NoStop}%
\bibitem [{\citenamefont {Yoshida}\ \emph {et~al.}(2022)\citenamefont {Yoshida}, \citenamefont {Mizoguchi},\ and\ \citenamefont {Hatsugai}}]{YMH2022}%
  \BibitemOpen
  \bibfield  {author} {\bibinfo {author} {\bibfnamefont {T.}~\bibnamefont {Yoshida}}, \bibinfo {author} {\bibfnamefont {T.}~\bibnamefont {Mizoguchi}},\ and\ \bibinfo {author} {\bibfnamefont {Y.}~\bibnamefont {Hatsugai}},\ }\bibfield  {title} {\bibinfo {title} {Non-hermitian topology in rock-paper-scissors games},\ }\href {https://doi.org/10.1038/s41598-021-04178-8} {\bibfield  {journal} {\bibinfo  {journal} {Sci. Rep.}\ }\textbf {\bibinfo {volume} {12}},\ \bibinfo {pages} {560} (\bibinfo {year} {2022})}\BibitemShut {NoStop}%
\bibitem [{\citenamefont {Knebel}\ \emph {et~al.}(2020)\citenamefont {Knebel}, \citenamefont {Geiger},\ and\ \citenamefont {Frey}}]{KGF2020}%
  \BibitemOpen
  \bibfield  {author} {\bibinfo {author} {\bibfnamefont {J.}~\bibnamefont {Knebel}}, \bibinfo {author} {\bibfnamefont {P.}~\bibnamefont {Geiger}},\ and\ \bibinfo {author} {\bibfnamefont {E.}~\bibnamefont {Frey}},\ }\bibfield  {title} {\bibinfo {title} {Topological phase transition in coupled rock-paper-scissors cycles},\ }\href {https://doi.org/10.1103/PhysRevLett.125.258301} {\bibfield  {journal} {\bibinfo  {journal} {Phys. Rev. Lett.}\ }\textbf {\bibinfo {volume} {125}},\ \bibinfo {pages} {258301} (\bibinfo {year} {2020})}\BibitemShut {NoStop}%
\bibitem [{\citenamefont {Liang}\ \emph {et~al.}(2024)\citenamefont {Liang}, \citenamefont {Dai}, \citenamefont {Li}, \citenamefont {Li},\ and\ \citenamefont {Yang}}]{LDL2024}%
  \BibitemOpen
  \bibfield  {author} {\bibinfo {author} {\bibfnamefont {J.}~\bibnamefont {Liang}}, \bibinfo {author} {\bibfnamefont {Q.}~\bibnamefont {Dai}}, \bibinfo {author} {\bibfnamefont {H.}~\bibnamefont {Li}}, \bibinfo {author} {\bibfnamefont {H.}~\bibnamefont {Li}},\ and\ \bibinfo {author} {\bibfnamefont {J.}~\bibnamefont {Yang}},\ }\bibfield  {title} {\bibinfo {title} {Topological phases in population dynamics with rock-paper-scissors interactions},\ }\href {https://doi.org/10.1103/PhysRevE.110.034208} {\bibfield  {journal} {\bibinfo  {journal} {Phys. Rev. E}\ }\textbf {\bibinfo {volume} {110}},\ \bibinfo {pages} {034208} (\bibinfo {year} {2024})}\BibitemShut {NoStop}%
\bibitem [{\citenamefont {Zhang}\ and\ \citenamefont {Cai}(2023)}]{ZC2023}%
  \BibitemOpen
  \bibfield  {author} {\bibinfo {author} {\bibfnamefont {T.}~\bibnamefont {Zhang}}\ and\ \bibinfo {author} {\bibfnamefont {Z.}~\bibnamefont {Cai}},\ }\bibfield  {title} {\bibinfo {title} {Emergent non-hermitian physics in a generalized lotka-volterra model},\ }\href {https://doi.org/10.1103/PhysRevB.108.104304} {\bibfield  {journal} {\bibinfo  {journal} {Phys. Rev. B}\ }\textbf {\bibinfo {volume} {108}},\ \bibinfo {pages} {104304} (\bibinfo {year} {2023})}\BibitemShut {NoStop}%
\bibitem [{\citenamefont {Hofbauer}\ and\ \citenamefont {Sigmund}(1998)}]{HS1998}%
  \BibitemOpen
  \bibfield  {author} {\bibinfo {author} {\bibfnamefont {J.}~\bibnamefont {Hofbauer}}\ and\ \bibinfo {author} {\bibfnamefont {K.}~\bibnamefont {Sigmund}},\ }\href {https://doi.org/10.1017/CBO9781139173179} {\emph {\bibinfo {title} {Evolutionary Games and Population Dynamics}}}\ (\bibinfo  {publisher} {Cambridge University Press},\ \bibinfo {year} {1998})\BibitemShut {NoStop}%
\bibitem [{Note1()}]{Note1}%
  \BibitemOpen
  \bibinfo {note} {In general, winning players spawn duplicate players following their strategy according to some growth rate, while loosing players are culled according to some decay rate.}\BibitemShut {Stop}%
\bibitem [{\citenamefont {Taylor}\ and\ \citenamefont {Jonker}(1978)}]{TY1978}%
  \BibitemOpen
  \bibfield  {author} {\bibinfo {author} {\bibfnamefont {P.~D.}\ \bibnamefont {Taylor}}\ and\ \bibinfo {author} {\bibfnamefont {L.~B.}\ \bibnamefont {Jonker}},\ }\bibfield  {title} {\bibinfo {title} {Evolutionary stable strategies and game dynamics},\ }\href {https://doi.org/https://doi.org/10.1016/0025-5564(78)90077-9} {\bibfield  {journal} {\bibinfo  {journal} {Math. Biosci.}\ }\textbf {\bibinfo {volume} {40}},\ \bibinfo {pages} {145} (\bibinfo {year} {1978})}\BibitemShut {NoStop}%
\bibitem [{\citenamefont {Hofbauer}\ \emph {et~al.}(1979)\citenamefont {Hofbauer}, \citenamefont {Schuster},\ and\ \citenamefont {Sigmund}}]{HSS1979}%
  \BibitemOpen
  \bibfield  {author} {\bibinfo {author} {\bibfnamefont {J.}~\bibnamefont {Hofbauer}}, \bibinfo {author} {\bibfnamefont {P.}~\bibnamefont {Schuster}},\ and\ \bibinfo {author} {\bibfnamefont {K.}~\bibnamefont {Sigmund}},\ }\bibfield  {title} {\bibinfo {title} {A note on evolutionary stable strategies and game dynamics},\ }\href {https://doi.org/https://doi.org/10.1016/0022-5193(79)90058-4} {\bibfield  {journal} {\bibinfo  {journal} {J. Theor. Biol.}\ }\textbf {\bibinfo {volume} {81}},\ \bibinfo {pages} {609} (\bibinfo {year} {1979})}\BibitemShut {NoStop}%
\bibitem [{\citenamefont {Szolnoki}\ \emph {et~al.}(2014)\citenamefont {Szolnoki}, \citenamefont {Mobilia}, \citenamefont {Jiang}, \citenamefont {Szczesny}, \citenamefont {Rucklidge},\ and\ \citenamefont {Perc}}]{SMJ2014}%
  \BibitemOpen
  \bibfield  {author} {\bibinfo {author} {\bibfnamefont {A.}~\bibnamefont {Szolnoki}}, \bibinfo {author} {\bibfnamefont {M.}~\bibnamefont {Mobilia}}, \bibinfo {author} {\bibfnamefont {L.-L.}\ \bibnamefont {Jiang}}, \bibinfo {author} {\bibfnamefont {B.}~\bibnamefont {Szczesny}}, \bibinfo {author} {\bibfnamefont {A.}~\bibnamefont {Rucklidge}},\ and\ \bibinfo {author} {\bibfnamefont {M.}~\bibnamefont {Perc}},\ }\bibfield  {title} {\bibinfo {title} {Cyclic dominance in evolutionary games: A review},\ }\href {https://doi.org/10.1098/rsif.2014.0735} {\bibfield  {journal} {\bibinfo  {journal} {J. R. Soc. Interface}\ }\textbf {\bibinfo {volume} {11}},\ \bibinfo {pages} {20140735} (\bibinfo {year} {2014})}\BibitemShut {NoStop}%
\bibitem [{\citenamefont {Kirillov}(2013)}]{K2013}%
  \BibitemOpen
  \bibfield  {author} {\bibinfo {author} {\bibfnamefont {O.~N.}\ \bibnamefont {Kirillov}},\ }\href {https://doi.org/https: //doi.org/10.1515/9783110270433} {\emph {\bibinfo {title} {Nonconservative Stability Problems of Modern Physics}}}\ (\bibinfo  {publisher} {Berlin, Boston: De Gruyter},\ \bibinfo {year} {2013})\BibitemShut {NoStop}%
\bibitem [{\citenamefont {Wiggins}(2010)}]{W2010}%
  \BibitemOpen
  \bibfield  {author} {\bibinfo {author} {\bibfnamefont {S.}~\bibnamefont {Wiggins}},\ }\href {https://doi.org/https://doi.org/10.1007/b97481} {\emph {\bibinfo {title} {Introduction to Applied Nonlinear Dynamical Systems and Chaos}}}\ (\bibinfo  {publisher} {Springer New York, NY},\ \bibinfo {year} {2010})\BibitemShut {NoStop}%
\bibitem [{\citenamefont {Bender}(2007)}]{B2007}%
  \BibitemOpen
  \bibfield  {author} {\bibinfo {author} {\bibfnamefont {C.~M.}\ \bibnamefont {Bender}},\ }\bibfield  {title} {\bibinfo {title} {Making sense of non-hermitian hamiltonians},\ }\href {https://doi.org/10.1088/0034-4885/70/6/R03} {\bibfield  {journal} {\bibinfo  {journal} {Rep. Prog. Phys.}\ }\textbf {\bibinfo {volume} {70}},\ \bibinfo {pages} {947} (\bibinfo {year} {2007})}\BibitemShut {NoStop}%
\bibitem [{\citenamefont {Brody}(2013)}]{B2013}%
  \BibitemOpen
  \bibfield  {author} {\bibinfo {author} {\bibfnamefont {D.}~\bibnamefont {Brody}},\ }\bibfield  {title} {\bibinfo {title} {Biorthogonal quantum mechanics},\ }\href {https://doi.org/10.1088/1751-8113/47/3/035305} {\bibfield  {journal} {\bibinfo  {journal} {J. Phys. A: Math. Theor.}\ }\textbf {\bibinfo {volume} {47}},\ \bibinfo {pages} {035305} (\bibinfo {year} {2013})}\BibitemShut {NoStop}%
\bibitem [{Note2()}]{Note2}%
  \BibitemOpen
  \bibinfo {note} {The model here differs slightly from \cite {YMH2022} for clarity, but follows their discussion in keeping with a coexistence point at equidistributed strategies.}\BibitemShut {Stop}%
\bibitem [{\citenamefont {Guckenheimer}\ and\ \citenamefont {Holmes}(1983)}]{GH1983}%
  \BibitemOpen
  \bibfield  {author} {\bibinfo {author} {\bibfnamefont {J.}~\bibnamefont {Guckenheimer}}\ and\ \bibinfo {author} {\bibfnamefont {P.}~\bibnamefont {Holmes}},\ }\href {https://doi.org/https://doi.org/10.1007/978-1-4612-1140-2} {\emph {\bibinfo {title} {Nonlinear Oscillations, Dynamical Systems, and Bifurcations of Vector Fields}}}\ (\bibinfo  {publisher} {Springer New York, NY},\ \bibinfo {year} {1983})\BibitemShut {NoStop}%
\bibitem [{\citenamefont {Hill}\ \emph {et~al.}(2024)\citenamefont {Hill}, \citenamefont {Gohsrich}, \citenamefont {Fauman}, \citenamefont {Ghosh}, \citenamefont {Kawagoe}, \citenamefont {Del'Haye},\ and\ \citenamefont {Kunst}}]{HGF2024}%
  \BibitemOpen
  \bibfield  {author} {\bibinfo {author} {\bibfnamefont {L.}~\bibnamefont {Hill}}, \bibinfo {author} {\bibfnamefont {L.}~\bibnamefont {Gohsrich}}, \bibinfo {author} {\bibfnamefont {J.}~\bibnamefont {Fauman}}, \bibinfo {author} {\bibfnamefont {A.}~\bibnamefont {Ghosh}}, \bibinfo {author} {\bibfnamefont {K.}~\bibnamefont {Kawagoe}}, \bibinfo {author} {\bibfnamefont {P.}~\bibnamefont {Del'Haye}},\ and\ \bibinfo {author} {\bibfnamefont {F.~K.}\ \bibnamefont {Kunst}},\ }\bibfield  {title} {\bibinfo {title} {Separating spontaneous symmetry breaking from exceptional points},\ }in\ \href {https://doi.org/10.1364/NP.2024.NpTh2D.4} {\emph {\bibinfo {booktitle} {Advanced Photonics Congress 2024}}}\ (\bibinfo  {publisher} {Optica Publishing Group},\ \bibinfo {year} {2024})\ p.\ \bibinfo {pages} {NpTh2D.4}\BibitemShut {NoStop}%
\bibitem [{\citenamefont {Feng}\ \emph {et~al.}(2012)\citenamefont {Feng}, \citenamefont {Xu}, \citenamefont {Fegadolli}, \citenamefont {Lu}, \citenamefont {Oliveira}, \citenamefont {Almeida}, \citenamefont {Chen},\ and\ \citenamefont {Scherer}}]{FXF2012}%
  \BibitemOpen
  \bibfield  {author} {\bibinfo {author} {\bibfnamefont {L.}~\bibnamefont {Feng}}, \bibinfo {author} {\bibfnamefont {Y.-L.}\ \bibnamefont {Xu}}, \bibinfo {author} {\bibfnamefont {W.}~\bibnamefont {Fegadolli}}, \bibinfo {author} {\bibfnamefont {M.-H.}\ \bibnamefont {Lu}}, \bibinfo {author} {\bibfnamefont {J.}~\bibnamefont {Oliveira}}, \bibinfo {author} {\bibfnamefont {V.}~\bibnamefont {Almeida}}, \bibinfo {author} {\bibfnamefont {Y.-F.}\ \bibnamefont {Chen}},\ and\ \bibinfo {author} {\bibfnamefont {A.}~\bibnamefont {Scherer}},\ }\bibfield  {title} {\bibinfo {title} {Experimental demonstration of a unidirectional reflectionless parity-time metamaterial at optical frequencies},\ }\href {https://doi.org/10.1038/nmat3495} {\bibfield  {journal} {\bibinfo  {journal} {Nature Mater}\ }\textbf {\bibinfo {volume} {12}},\ \bibinfo {pages} {108} (\bibinfo {year} {2012})}\BibitemShut {NoStop}%
\bibitem [{Note3()}]{Note3}%
  \BibitemOpen
  \bibinfo {note} {Note that this differs from the studies \cite {HDL2014, GKN2010} of local and global stability in $\protect \ensuremath {\protect \mathcal {PT}}$-symmetric nonlinear systems, wherein the behavior of the position of stationary states under $\protect \ensuremath {\protect \mathcal {PT}}$ transformation is considered in a globally symmetric model. In contrast, the spontaneous symmetry breaking transition, marked by an EP, arises in the local stability characterization of an isolated fixed-position equilibrium in this study. Moreover, the connection of global and local stability is analyzed for globally symmetric systems, as well as for systems in which such a symmetry is absent}\BibitemShut {NoStop}%
\bibitem [{\citenamefont {Delplace}\ \emph {et~al.}(2021)\citenamefont {Delplace}, \citenamefont {Yoshida},\ and\ \citenamefont {Hatsugai}}]{DJH2021}%
  \BibitemOpen
  \bibfield  {author} {\bibinfo {author} {\bibfnamefont {P.}~\bibnamefont {Delplace}}, \bibinfo {author} {\bibfnamefont {T.}~\bibnamefont {Yoshida}},\ and\ \bibinfo {author} {\bibfnamefont {Y.}~\bibnamefont {Hatsugai}},\ }\bibfield  {title} {\bibinfo {title} {Symmetry-protected multifold exceptional points and their topological characterization},\ }\href {https://doi.org/10.1103/PhysRevLett.127.186602} {\bibfield  {journal} {\bibinfo  {journal} {Phys. Rev. Lett.}\ }\textbf {\bibinfo {volume} {127}},\ \bibinfo {pages} {186602} (\bibinfo {year} {2021})}\BibitemShut {NoStop}%
\bibitem [{\citenamefont {Mandal}\ and\ \citenamefont {Bergholtz}(2021)}]{MB2021}%
  \BibitemOpen
  \bibfield  {author} {\bibinfo {author} {\bibfnamefont {I.}~\bibnamefont {Mandal}}\ and\ \bibinfo {author} {\bibfnamefont {E.~J.}\ \bibnamefont {Bergholtz}},\ }\bibfield  {title} {\bibinfo {title} {Symmetry and higher-order exceptional points},\ }\href {https://doi.org/10.1103/PhysRevLett.127.186601} {\bibfield  {journal} {\bibinfo  {journal} {Phys. Rev. Lett.}\ }\textbf {\bibinfo {volume} {127}},\ \bibinfo {pages} {186601} (\bibinfo {year} {2021})}\BibitemShut {NoStop}%
\bibitem [{\citenamefont {Lotka}(1920)}]{L1920}%
  \BibitemOpen
  \bibfield  {author} {\bibinfo {author} {\bibfnamefont {A.}~\bibnamefont {Lotka}},\ }\bibfield  {title} {\bibinfo {title} {Analytical note on certain rhythmic relations in organic systems},\ }\href {https://doi.org/10.1073/pnas.6.7.410} {\bibfield  {journal} {\bibinfo  {journal} {Proc. Natl. Acad. Sci. U.S.A.}\ }\textbf {\bibinfo {volume} {6}},\ \bibinfo {pages} {410} (\bibinfo {year} {1920})}\BibitemShut {NoStop}%
\bibitem [{\citenamefont {Volterra}(1926)}]{V1926}%
  \BibitemOpen
  \bibfield  {author} {\bibinfo {author} {\bibfnamefont {V.}~\bibnamefont {Volterra}},\ }\bibfield  {title} {\bibinfo {title} {Fluctuations in the abundance of a species considered mathematically},\ }\href {https://doi.org/10.1038/118558a0} {\bibfield  {journal} {\bibinfo  {journal} {Nature}\ }\textbf {\bibinfo {volume} {118}},\ \bibinfo {pages} {558} (\bibinfo {year} {1926})}\BibitemShut {NoStop}%
\bibitem [{\citenamefont {Yi}(2010)}]{Y2010}%
  \BibitemOpen
  \bibfield  {author} {\bibinfo {author} {\bibfnamefont {Z.}~\bibnamefont {Yi}},\ }\bibfield  {title} {\bibinfo {title} {Foundations of implementing the competitive layer model by lotka-volterra recurrent neural networks},\ }\href {https://doi.org/10.1109/TNN.2009.2039758} {\bibfield  {journal} {\bibinfo  {journal} {IEEE Trans. Neural Netw.}\ }\textbf {\bibinfo {volume} {21}},\ \bibinfo {pages} {494} (\bibinfo {year} {2010})}\BibitemShut {NoStop}%
\bibitem [{\citenamefont {Laval}\ and\ \citenamefont {Pellat}(1975)}]{LP1975}%
  \BibitemOpen
  \bibfield  {author} {\bibinfo {author} {\bibfnamefont {G.}~\bibnamefont {Laval}}\ and\ \bibinfo {author} {\bibfnamefont {R.}~\bibnamefont {Pellat}},\ }\bibfield  {title} {\bibinfo {title} {Plasma physics},\ }in\ \href@noop {} {\emph {\bibinfo {booktitle} {Proceedings of Summer School of Theoretical Physics}}}\ (\bibinfo  {publisher} {Gordon and Breach, New York},\ \bibinfo {year} {1975})\BibitemShut {NoStop}%
\bibitem [{\citenamefont {Orlando}\ and\ \citenamefont {Sportelli}(2021)}]{OS2021}%
  \BibitemOpen
  \bibfield  {author} {\bibinfo {author} {\bibfnamefont {G.}~\bibnamefont {Orlando}}\ and\ \bibinfo {author} {\bibfnamefont {M.}~\bibnamefont {Sportelli}},\ }\bibinfo {title} {Growth and cycles as a struggle: Lotka-volterra, goodwin and phillips}\ (\bibinfo {year} {2021})\ pp.\ \bibinfo {pages} {191--208}\BibitemShut {NoStop}%
\bibitem [{\citenamefont {Kon}(2004)}]{K2004}%
  \BibitemOpen
  \bibfield  {author} {\bibinfo {author} {\bibfnamefont {R.}~\bibnamefont {Kon}},\ }\bibfield  {title} {\bibinfo {title} {A note on constants of motion for the lotka-volterra and replicator equations},\ }in\ \href@noop {} {\emph {\bibinfo {booktitle} {Hyperbolic Problems, Theory, Numerics and Applications II}}}\ (\bibinfo  {publisher} {Yokohama Publishers},\ \bibinfo {year} {2004})\ pp.\ \bibinfo {pages} {109--116}\BibitemShut {NoStop}%
\bibitem [{\citenamefont {Imane}\ \emph {et~al.}(2024)\citenamefont {Imane}, \citenamefont {Matthieu}, \citenamefont {Maxime}, \citenamefont {Mylene}, \citenamefont {Francois}, \citenamefont {Jamal},\ and\ \citenamefont {Chi}}]{ABC2024}%
  \BibitemOpen
  \bibfield  {author} {\bibinfo {author} {\bibfnamefont {A.}~\bibnamefont {Imane}}, \bibinfo {author} {\bibfnamefont {B.}~\bibnamefont {Matthieu}}, \bibinfo {author} {\bibfnamefont {H.}~\bibnamefont {Maxime}, \bibfnamefont {C.and~Walid}}, \bibinfo {author} {\bibfnamefont {M.}~\bibnamefont {Mylene}}, \bibinfo {author} {\bibfnamefont {M.}~\bibnamefont {Francois}}, \bibinfo {author} {\bibfnamefont {N.}~\bibnamefont {Jamal}},\ and\ \bibinfo {author} {\bibfnamefont {T.~V.}\ \bibnamefont {Chi}},\ }\bibfield  {title} {\bibinfo {title} {Complex systems in ecology: a guided tour with large lotka-volterra models and random matrices},\ }\href {https://doi.org/10.1098/rspa.2023.0284} {\bibfield  {journal} {\bibinfo  {journal} {Proc. R. Soc. A.}\ }\textbf {\bibinfo {volume} {480}},\ \bibinfo {pages} {20230284} (\bibinfo {year} {2024})}\BibitemShut {NoStop}%
\bibitem [{\citenamefont {Busiello}\ \emph {et~al.}(2017)\citenamefont {Busiello}, \citenamefont {Suweis}, \citenamefont {Hidalgo},\ and\ \citenamefont {Maritan}}]{BSH2017}%
  \BibitemOpen
  \bibfield  {author} {\bibinfo {author} {\bibfnamefont {D.~M.}\ \bibnamefont {Busiello}}, \bibinfo {author} {\bibfnamefont {S.}~\bibnamefont {Suweis}}, \bibinfo {author} {\bibfnamefont {J.}~\bibnamefont {Hidalgo}},\ and\ \bibinfo {author} {\bibfnamefont {A.}~\bibnamefont {Maritan}},\ }\bibfield  {title} {\bibinfo {title} {Explorability and the origin of network sparsity in living systems},\ }\href {https://doi.org/10.1038/s41598-017-12521-1} {\bibfield  {journal} {\bibinfo  {journal} {Sci. Rep.}\ }\textbf {\bibinfo {volume} {7}},\ \bibinfo {pages} {12323} (\bibinfo {year} {2017})}\BibitemShut {NoStop}%
\bibitem [{\citenamefont {Yodzis}(2001)}]{Y2001}%
  \BibitemOpen
  \bibfield  {author} {\bibinfo {author} {\bibfnamefont {P.}~\bibnamefont {Yodzis}},\ }\bibfield  {title} {\bibinfo {title} {Trophic levels},\ }in\ \href {https://doi.org/10.1016/B978-0-12-384719-5.00145-3} {\emph {\bibinfo {booktitle} {Encyclopedia of Biodiversity (Second Edition)}}}\ (\bibinfo  {publisher} {Academic Press},\ \bibinfo {year} {2001})\ pp.\ \bibinfo {pages} {264--268}\BibitemShut {NoStop}%
\bibitem [{\citenamefont {Abdala-Roberts}\ \emph {et~al.}(2019)\citenamefont {Abdala-Roberts}, \citenamefont {Puentes}, \citenamefont {Finke}, \citenamefont {Marquis}, \citenamefont {Montserrat}, \citenamefont {Poelman}, \citenamefont {Rasmann}, \citenamefont {Sentis}, \citenamefont {Dam}, \citenamefont {Wimp}, \citenamefont {Mooney},\ and\ \citenamefont {Bj\"orkman}}]{APF2019}%
  \BibitemOpen
  \bibfield  {author} {\bibinfo {author} {\bibfnamefont {L.}~\bibnamefont {Abdala-Roberts}}, \bibinfo {author} {\bibfnamefont {A.}~\bibnamefont {Puentes}}, \bibinfo {author} {\bibfnamefont {D.}~\bibnamefont {Finke}}, \bibinfo {author} {\bibfnamefont {R.}~\bibnamefont {Marquis}}, \bibinfo {author} {\bibfnamefont {M.}~\bibnamefont {Montserrat}}, \bibinfo {author} {\bibfnamefont {E.}~\bibnamefont {Poelman}}, \bibinfo {author} {\bibfnamefont {S.}~\bibnamefont {Rasmann}}, \bibinfo {author} {\bibfnamefont {A.}~\bibnamefont {Sentis}}, \bibinfo {author} {\bibfnamefont {N.}~\bibnamefont {Dam}}, \bibinfo {author} {\bibfnamefont {G.}~\bibnamefont {Wimp}}, \bibinfo {author} {\bibfnamefont {K.}~\bibnamefont {Mooney}},\ and\ \bibinfo {author} {\bibfnamefont {C.}~\bibnamefont {Bj\"orkman}},\ }\bibfield  {title} {\bibinfo {title} {Tri-trophic interactions: bridging species, communities and ecosystems},\ }\href {https://doi.org/10.1111/ele.13392} {\bibfield  {journal} {\bibinfo  {journal} {Ecol. Lett.}\ }\textbf {\bibinfo
  {volume} {22}},\ \bibinfo {pages} {2151} (\bibinfo {year} {2019})}\BibitemShut {NoStop}%
\bibitem [{\citenamefont {Liu}\ \emph {et~al.}(2009)\citenamefont {Liu}, \citenamefont {Liu}, \citenamefont {An}, \citenamefont {Fu}, \citenamefont {Zeng}, \citenamefont {Luo}, \citenamefont {Wu},\ and\ \citenamefont {Pratley}}]{LLA2009}%
  \BibitemOpen
  \bibfield  {author} {\bibinfo {author} {\bibfnamefont {Y.}~\bibnamefont {Liu}}, \bibinfo {author} {\bibfnamefont {D.~L.}\ \bibnamefont {Liu}}, \bibinfo {author} {\bibfnamefont {M.}~\bibnamefont {An}}, \bibinfo {author} {\bibfnamefont {Y.}~\bibnamefont {Fu}}, \bibinfo {author} {\bibfnamefont {R.~S.}\ \bibnamefont {Zeng}}, \bibinfo {author} {\bibfnamefont {S.~M.}\ \bibnamefont {Luo}}, \bibinfo {author} {\bibfnamefont {H.}~\bibnamefont {Wu}},\ and\ \bibinfo {author} {\bibfnamefont {J.}~\bibnamefont {Pratley}},\ }\bibfield  {title} {\bibinfo {title} {Modelling tritrophic interactions mediated by induced defence volatiles, ecological modelling},\ }\href {https://doi.org/10.1016/j.ecolmodel.2009.07.003} {\bibfield  {journal} {\bibinfo  {journal} {Ecol. Model.}\ }\textbf {\bibinfo {volume} {220}},\ \bibinfo {pages} {3241} (\bibinfo {year} {2009})}\BibitemShut {NoStop}%
\end{thebibliography}%

\end{document}